\documentclass[a4paper,12pt]{article}
\usepackage{jheppub,esint,shuffle,psfrag}
\usepackage[utf8]{inputenc}
\usepackage{booktabs}
\usepackage{breqn}
\usepackage{amsmath,bm}
\usepackage{hyperref}
\usepackage{mathrsfs} 
\usepackage{enumerate}
\usepackage{tikz-cd}
\usepackage{shuffle}
\usepackage{array}
\usepackage{tcolorbox}
\tcbuselibrary{hooks}
\tcbset{colback=black!3!white}
\makeatletter
\tcbset{
  after app={%
    \ifx\tcb@drawcolorbox\tcb@drawcolorbox@breakable
    \else
      \@endparenv
    \fi
  }
}
\makeatother
\usepackage{tikz}

\definecolor{myred}{RGB}{233, 33, 45}

\DeclareMathOperator{\tr}{\text{tr}}     
\newcommand{\vev}[1]{\big\langle #1 \big\rangle}
\newcommand\re[1]{(\ref{#1})}

\newcommand\lr[1]{{\left({#1}\right)}}

\newcommand \ket [1] {|{#1}\rangle}
\newcommand \bra [1] {\langle {#1}|}
\def \qqquad {\qquad\quad}
\def \qqqquad {\qquad\qquad}

\newcommand{\ft}[2]{{\textstyle\frac{#1}{#2}}}
 \def\numberbysection{\@addtoreset{equation}{section}
                     \def\theequation{\thesection.\arabic{equation}}}                     
 
\vspace*{5mm}      

\title{Lattice path combinatorics in superconformal Yang-Mills theories}

\author{G.P. Korchemsky}  
\affiliation{Institut de Physique Th\'eorique\footnote{Unit\'e Mixte de Recherche 3681 du CNRS}, Universit\'e Paris Saclay, CNRS,  91191 Gif-sur-Yvette, France} 

\abstract{We study a class of observables in four-dimensional superconformal Yang--Mills theories which, in the planar limit at finite 't Hooft coupling, can be expressed as determinants of semi-infinite matrices built from Bessel functions. This determinant representation points to an underlying integrable structure, which we make explicit by showing that the observables satisfy a nonlinear differential-difference equation. We argue that the solution to this equation admits an expansion in terms of iterated Chen integrals of uniform transcendental weight. Remarkably, the coefficients in this expansion are universal positive integers, independent of the particular observable, suggesting a hidden combinatorial origin. Building on this observation, we show that the resulting expressions possess a natural interpretation in enumerative combinatorics: they coincide with the partition function (or generating function) of an ensemble of lattice paths constrained to a nontrivial domain. This correspondence extends and generalizes the classical Dyck paths to a richer family of path ensembles relevant in gauge theory.
}

\begin{document}
\maketitle

\section{Introduction and summary}

In this paper, we continue the study~\cite{Bajnok:2024epf,Bajnok:2024ymr,Bajnok:2024bqr} of a broad class of observables in four-dimensional superconformal Yang--Mills theories that can be computed in the planar limit at finite ’t Hooft coupling $\lambda$. A distinguishing feature of these observables is that, for arbitrary values of the coupling $g = \sqrt{\lambda}/(4\pi)$, they admit a representation in terms of determinants of certain semi-infinite Bessel matrices
\begin{align}\label{F(g)}
D_\ell(g) = \det\left(\delta_{nm}-K_{nm}(g)\right)\Big|_{1\le n,m <\infty}\,.
\end{align}
In addition to the coupling constant, it depends on a real parameter $\ell$ and a real-valued function $\chi(x)$ that are specified below. Our goal is to demonstrate that the determinant \re{F(g)} can be naturally reinterpreted as the partition function of an ensemble of lattice paths, thereby uncovering its combinatorial meaning.

The representation \re{F(g)} was derived across a range of gauge theories and for various observables, see e.g. \cite{Beisert:2006ez,Kostov:2019stn,Kostov:2019auq,Belitsky:2019fan,Basso:2020xts,Belitsky:2020qrm,Belitsky:2020qir,Beccaria:2020hgy,Beccaria:2021vuc,Beccaria:2021hvt,Billo:2021rdb,Beccaria:2022ypy,Beccaria:2023kbl}, utilizing different methods such as integrability \cite{Beisert:2010jr} and localization \cite{Pestun_2017}. It is therefore remarkable that, despite these diverse settings, the matrix in \re{F(g)} takes a universal form
\begin{align} \label{eq:K_nm}
K_{nm}(g)=\int_{0}^{\infty}dx\,\psi_{n}(x)\chi\Big(\frac{\sqrt{x}}{2g}\Big)\psi_{m}(x)\,,
\end{align}
where $\psi_n(x)$ are given by the normalized Bessel functions (hence the name of the matrix)
\begin{equation}\label{psi}
\psi_{n}(x)=\sqrt{2n+\ell-1}{J_{2n+\ell-1}(\sqrt{x})\over\sqrt{x}}\,.
\end{equation}
In this relation, the argument of the Bessel function and the normalization factor are chosen to simplify the orthogonality condition $\int_{0}^{\infty}dx\,\psi_{n}(x)\psi_{m}(x)=\delta_{nm}$. 
The function \re{F(g)} is known in the mathematical literature as a Fredholm determinant of the truncated Bessel operator~\cite{BasorEhrhardt03}. 

The dependence of \re{F(g)} and \re{eq:K_nm} on the coupling constant $g$ enters through the argument of the function $\chi(x)$, conventionally called the symbol of the matrix \re{eq:K_nm}.  
The dependence  of \re{F(g)} on the real parameter $\ell$ resides in the function \re{psi}. Assuming that the symbol function behaves at the origin as  $\chi(x)\sim x^{2\beta}$, the condition for 
the integral in \re{eq:K_nm} to be convergent at $x=0$ leads to the condition
\begin{align}\label{ell-cond}
\ell > -1-\beta\,.
\end{align}
Additional simplification occurs when $\ell$ takes positive integer values. In this case, depending on the parity of $\ell$, the matrix \re{eq:K_nm} can be obtained from the same matrix evaluated at $\ell=0$ or $\ell=1$ by removing the first $\ell/2$ and 
$(\ell-1)/2$ rows and columns for even and odd $\ell$, respectively. 

In the context of four-dimensional supersymmetric gauge theories, we encounter various types of the symbol function
\begin{align}\notag\label{chis}
& \chi_{\text{W}}(x) = - {(2\pi)^2\over x^2}\,,  
\\[2mm]\notag
& \chi_{\text{ft}}(x) = 1- \coth(x/2)\,, 
\\[2mm]\notag
& \chi_{\text{loc}}(x) =- {1\over\sinh^{2}(x/2)} \,,
\\ 
& \chi_{\text{oct}}(x|y,\xi) = {\cosh y + \cosh \xi\over \cosh y + \cosh \sqrt{x^2+\xi^2}}\,.
\end{align}
In the last relation, $y$ and $\xi$ are the auxiliary (kinematical) variables.
Here the subscript refers to the observable it describes -- the circular Wilson loop~\cite{Semenoff:2001xp,Beccaria:2023kbl}, flux tube correlators~\cite{Beisert:2006ez,Belitsky:2019fan,Basso:2020xts}, localization integrals~\cite{Beccaria:2020hgy,Beccaria:2021vuc,Beccaria:2021hvt,Billo:2021rdb,Beccaria:2022ypy,Beccaria:2023kbl} and the octagon form factor~\cite{Coronado:2018ypq,Coronado:2018cxj,Kostov:2019stn,Kostov:2019auq,Belitsky:2020qrm,Belitsky:2020qir}.\,\footnote{We refer the interested reader to \cite{Bajnok:2024ymr} for more details.} 
The first function in \re{chis} 
has a power-like behaviour at infinity, $\chi(x) \sim 1/x^2$, while the three remaining functions decay exponentially fast $\chi(x) \sim e^{-x}$. In application to gauge theories, the parameter $\ell$ in \re{F(g)} takes nonnegative integer values. Additionally, 
due to \re{ell-cond}, for the first and third functions in \re{chis}, this parameter has to satisfy the condition $\ell\ge 1$.

The properties of the determinant \re{F(g)} crucially depend on the choice of the symbol function. 
In the special case $\chi(x) = 1$, the matrix in \re{eq:K_nm} reduces to the identity, leading to the trivial result $D_\ell(g) = 0$. In contrast, the functions in \re{chis} vanish as $x \to \infty$, effectively suppressing the contribution to the integral in \re{eq:K_nm} from the region $x \gg (2g)^2$. As a result, the corresponding functions $D_\ell(g)$ are nonzero and exhibit a nontrivial dependence on the coupling constant.

Another interesting case arises when the symbol function $\chi(x)$ takes the value 1 on a union of intervals $J = [a_1, b_1] \cup \dots \cup [a_m, b_m]$ on the positive semi-axis and vanishes elsewhere. In this setting, the determinant \re{F(g)} describes the eigenvalue spacing near the hard edge of the Laguerre unitary ensemble \cite{Forrester:1993vtx}, and is known to satisfy a system of integrable partial differential equations, see e.g. \cite{Korepin:1993kvr,Harnad_Balogh_2021}. For a single interval, with the symbol function given by
\begin{align}\label{chi-TW}
\chi_{\text{TW}}(x) = \theta(1 - x)\,,
\end{align}
the function \re{F(g)} coincides with the Tracy–Widom distribution~\cite{Tracy:1993xj}. It gives the probability that no eigenvalues lie in the interval $J = [0, s]$ for $s = (2g)^2$. For arbitrary $g$ and $\ell$, the corresponding function $D_\ell(g)$ can be computed explicitly in terms of solutions to the Painlev\'e~V equation~\cite{Okamoto,kajiwara,Witte}.
 
Computing the determinant \re{F(g)} for the symbol functions defined in \re{chis} is much more complicated than in the case of \re{chi-TW}. For a generic symbol function $\chi(x)$ in \re{chis}, the corresponding function $D_\ell(g)$ satisfies an integro-differential equation with respect to the coupling $g$ \cite{Belitsky:2019fan,Belitsky:2020qrm,Belitsky:2020qir}. For the function \re{chi-TW}, this equation reduces to the Painlev\'e~V differential equation.\,\footnote{The reason is that the derivative $\partial_x \chi_{\text{TW}}(x) = -\delta(1 - x)$ localizes the integrals in the equation at $x = 1$.} For the functions defined in \re{chis}, the integro-differential equation does not have a closest-form solution but it can be used to derive a systematic expansion of $D_\ell(g)$ at small and large $g$. These two expansions can be sewed together at finite $g$, enabling us to determine $D_\ell(g)$ over a wide range of the coupling. Remarkably,  for the first two functions in \re{chis}, the determinant $D_\ell(g)$ can be computed exactly in closed form for arbitrary $g$. The corresponding expressions are presented below (see \re{D-Bessel} and \re{Dft-spec}).

\subsection*{Differential-difference equation }

In addition to an integro-differential equation mentioned above, the determinant \re{F(g)} satisfies another differential-difference equation 
for arbitrary $\ell$ and $g$  
\begin{align}\label{Id}
g\partial_g \log \lr{D_{\ell+1}\over D_{\ell-1}} = 2\ell\left({D_\ell^2\over  D_{\ell-1}D_{\ell+1}} -1\right).
\end{align}
We would like to emphasize that this equation holds for arbitrary symbol function, including those defined in \re{chis} and \re{chi-TW}. For the function $\chi_{\text{loc}}(x)$ defined in \re{chis}, the relation \re{Id} was proposed in \cite{Ferrando:2025qkr,Korchemsky:2025eyc} in the context of four-dimensional superconformal $\mathcal N=2$ Yang-Mills theories, where it emerged from the study of correlation functions using two complementary approaches: integrability and localization. A derivation of the equation \re{Id} for a general symbol function is provided in Appendix~\ref{App:A}.

Equation \re{Id} proves to be very powerful: by applying it recursively, we can express \( D_{\ell+n}(g) \) for any positive integer \( n \) entirely in terms of the initial data \( D_{\ell-1}(g) \) and \( D_\ell(g) \). We show that the resulting expression admits the following schematic form
\begin{align}\label{D-schem}
D_{\ell+n}(g) \sim \sum c_{\sigma_1\sigma_2\dots\sigma_k} \, I_{\sigma_1\sigma_2\cdots\sigma_k}(g)\,,
\end{align}
where the sum runs over sequences \( (\sigma_1\sigma_2\dots\sigma_k) \) of signs \( \sigma_i = \pm \) of length \( k = n(n-1)/2 \). The functions \( I_{\sigma_1\sigma_2\cdots\sigma_k}(g) \) are iterated (Chen) integrals built from \( D_{\ell-1}(g) \) and \( D_\ell(g) \). Importantly, the expansion coefficients \( c_{\sigma_1\sigma_2\dots\sigma_k} \) are universal positive integers that depend only on \( n \),  and are completely independent of the parameters $\ell$, $g$ and 
the symbol function $\chi(x)$.

Iterated integrals are ubiquitous in quantum field theory, see e.g.~\cite{Vergu:2013,Abreu:2022mfk}. They arise naturally in the weak coupling expansion of observables, where their arguments are typically given by kinematical invariants. An unusual feature of relation \re{D-schem} is that the argument of the iterated integral depends on the coupling constant, whereas the dependence on the kinematic variables is contained solely within the initial conditions for equation \re{Id}.

The solution~\re{D-schem} admits a natural and insightful interpretation within the framework of enumerative combinatorics \cite{Krattenthaler}. Specifically, each sequence \( (\sigma_1 \sigma_2 \dots \sigma_k) \) appearing in the sum in~\re{D-schem} can be mapped bijectively to a unique path on a two-dimensional lattice. In this mapping, the entries \( \sigma_i \) encode discrete steps of the path, so that the entire sum enumerates all possible paths consistent with certain boundary conditions and constraints.

From this perspective, \( D_{\ell+n}(g) \) can be understood as the partition function (or equivalently, the generating function) for an ensemble of lattice paths confined to a nontrivial domain shown in Figure~\ref{area} below. These lattice paths can be viewed as a generalization of classical Dyck paths. While standard Dyck paths consist of up-steps and down-steps constrained to stay above the horizontal axis, the paths arising here allow for more general boundary conditions dictated by the recursive structure of equation \re{Id}.

This combinatorial viewpoint provides a natural explanation for the structure and properties of the expansion coefficients in~\re{D-schem}: each coefficient corresponds to the total weight of generalized Dyck paths satisfying these constraints. Moreover, it reveals a deep connection between the recursive structure of equation \re{Id} and classical counting problems in lattice path combinatorics.  
 
The paper is organized as follows. In Section~\ref{sect2}, we present explicit solutions to the equation~\re{Id} corresponding to the symbol functions defined in \re{chis} and \re{chi-TW}. In Section~\ref{sect3}, we construct the general solution to \re{Id} and show that it takes the expected form \re{D-schem}. An interpretation of the solution \re{D-schem} in terms of paths on the two-dimensional lattice is given in Section~\ref{sect:paths}. Technical details are collected in several appendices.
 
\section{Special solutions}\label{sect2}
 
For some symbol functions defined in \re{chis} and \re{chi-TW}, the determinant \re{F(g)} can be computed in a closed form.  
In this  section, we present the corresponding expressions and verify that they satisfy the equation \re{Id}.

\subsection*{Tracy-Widom distribution}

For the symbol function \re{chi-TW} the determinant \re{F(g)} coincides with the Tracy-Widom distribution~\cite{Tracy:1993xj}
\begin{align}\label{D-TW} 
D_{\rm TW,\ell}(g) = E_\ell (s)\,,
\end{align}
evaluated at $s=(2g)^2$. 
For $\ell >-1$ and arbitrary positive $s$, this distribution is given by the following expression 
\begin{align}\label{E} 
 E_\ell(s) = \exp\lr{-\frac14\int_0^s dt \, \log(s/t) q_\ell^2(t)}\,, 
\end{align}
where   the function $q_\ell=q_\ell(s)$ satisfies differential equation (with $q_\ell'=dq_\ell(s)/ds$)
\begin{align}
s(q_\ell^2-1)(sq_\ell')' = q_\ell(s q_\ell')^2 +\frac14(s-\ell^2) q_\ell+\frac14 s q_\ell^3 (q_\ell^2-2)\,,
\end{align}
supplemented by the boundary condition $q_\ell(s) \sim {(s/4)^{\ell/2}/ \Gamma(1+\ell)}$ for $s\to 0$.
 
The simplest way to check that the function \re{D-TW} satisfies the equation \re{Id} is to examine its asymptotic behaviour at small and large $g$. 
At small $g$, or equivalently for $s\to 0$, it is straightforward to expand \re{E} in powers of $s$ and verify \re{Id}. At large $g$, or equivalently $s\gg 1$,
the function $E_\ell(s)$ decreases exponentially fast and has an asymptotic expansion
\begin{align}\label{E-asym}
 E_\ell(s)=c_\ell {e^{-s/4+\ell\sqrt s}\over s^{\ell^2/4}} \left[1+ {\ell\over 8}s^{-1/2} + {9\ell^2\over 128} s^{-1} + \lr{{3\ell\over 128}+{51\ell^3\over 1024}} s^{-3/2} +O(s^{-2})\right]\,,
\end{align}
where $c_\ell =  {G(1+\ell)/ (2\pi)^{\ell/2}}$ is proportional to the Barnes $G-$function. Substituting \re{D-TW} and the asymptotic expansion \re{E-asym} into \re{Id}, we find  that the equation is indeed satisfied, provided the constant $c_\ell$
obeys the recurrence relation
\begin{align}
c_{\ell+1}c_{\ell-1} = \ell c_\ell^2 \,.
\end{align} 
This identity follows directly from the properties of the Barnes function.

\subsection*{Relation to the Toda equation}
 
The relations \re{D-TW} and \re{E} hold for arbitrary real $\ell>-1$. For nonnegative integer $\ell$, the the Tracy-Widom distribution $E_\ell(s)$ simplifies significantly. In this case, it admits another equivalent representation \cite{Forrester,*forrester2000}
\begin{align}\label{toda} 
E_{\ell}(s)= e^{-s/4} \tau_\ell(\sqrt s)\,,\qquad\qquad  \tau_\ell(s)=\det \Big[I_{j-k}( s) \Big]\Big|_{j,k=1,\dots,\ell}\,,
\end{align}
where $ \tau_0(s)=1$ and $I_n(s)$ is a modified Bessel function.  
The same relation \re{toda} also describes the determinant $D_{\text{oct},\ell}(g)$ corresponding to the last symbol in \re{chis} evaluated at the special kinematical point $y=\sqrt{s}/(2g)$ and $\xi=0$, in the limit $g\to 0$ (see \cite{Belitsky:2020qir}).
 
Using the properties of Bessel functions, one can show that the function $\tau_\ell=\tau_\ell(s)$ satisfies the following relations
\begin{align}\label{toda1}\notag
{}& (s\partial_s)^2 \log\tau_\ell = s^2 {\tau_{\ell-1} \tau_{\ell+1} \over \tau_\ell^2}\,,
\\
{}& s\partial_s \log \lr{\tau_{\ell+1}\over \tau_{\ell-1}} = 2\ell\left({\tau_\ell^2\over\tau_{\ell-1}\tau_{\ell+1}} -1\right).
\end{align}
The first relation can be recognized as a Toda lattice equation, see e.g. \cite{Harnad_Balogh_2021}. The second relation ensures that the function defined by \re{D-TW} and \re{toda} satisfies the equation \re{Id}. Although the right-hand sides of the two relations in \re{toda1} appear structurally similar, they are independent. In addition, they possess different symmetry --   the first relation in \re{toda1} is invariant under rescaling $\tau_\ell \to  s^{\ell} \tau_\ell$, while the second relation is  invariant under the transformation
$\tau_\ell \to f(s) \tau_\ell$.  
  
\subsection*{Circular Wilson loop}
 
For the symbol function $\chi_{\rm W}(x)$ defined in \re{chis}, the relation \re{ell-cond} leads to the condition $\ell>0$. In this case, in virtue of  the properties of the Bessel functions \re{psi}, the semi-infinite matrix \re{eq:K_nm} becomes tridiagonal. Diagonalizing this matrix, one can find its eigenspectrum and then evaluate its Fredholm determinant~\cite{Beccaria:2023kbl}  
\begin{align}\label{D-Bessel}
D_{\rm W,\ell}(g)= \Gamma(\ell) (2\pi g)^{1-\ell}  I_{\ell-1}(4\pi g) \,,
\end{align}
where $I_{\ell-1}(x)$ is a modified Bessel function. 

The relation \re{D-Bessel} holds for arbitrary $g$ and $\ell> 0$.  It is straightforward to verify that it satisfies the equation \re{Id}. At the same time, $D_{\rm W,\ell}(g)$ does not satisfy the Toda lattice equation. 
 
\subsection*{Flux tube correlators}

For the symbol function $\chi_{\rm ft}(x)$ in \re{chis}, the condition \re{ell-cond} leads to $\ell>-1/2$. 
The corresponding function \re{F(g)} can be computed exactly for the first few nonnegative integer values of $\ell$. Remarkably, for $\ell=0,1,2$ it can be expressed in terms of hyperbolic functions \cite{Beccaria:2022ypy}
\begin{align}\notag\label{Dft-spec}
{}&D_{\rm ft,\ell=0}(g)=  \left[\frac{2 \pi  g \cosh^3 (2 \pi  g)}{\sinh (2 \pi  g)}\right]^{1/8},
\\\notag
{}&D_{\rm ft,\ell=1}(g)=\left[\frac{\sinh^3(2 \pi  g)}{(2 \pi  g)^3 \cosh (2 \pi  g)}\right]^{1/8},
\\   
{}&D_{\rm ft,\ell=2}(g)=\frac{\log (\cosh (2 \pi  g))}{2 (\pi g)^2}\left[\frac{2 \pi  g \cosh^3 (2 \pi  g)}{\sinh (2 \pi  g)}\right]^{1/8} .
\end{align}
Certain combinations of these functions, $D_{\rm ft,\ell=2}/D_{\rm ft,\ell=0}$ and $D_{\rm ft,\ell=0}+D_{\rm ft,\ell=1}$, govern the asymptotic behaviour of the light-like octagon \cite{Belitsky:2019fan} and also appear in the analysis of scattering amplitudes in planar \(\mathcal{N}=4\) SYM at a special kinematical point \cite{Basso:2020xts}.
 
We can use the relations \re{Dft-spec} to verify the equation \re{Id} for $\ell=1$. However, as in the previous case, $D_{\rm ft,\ell}(g)$ does not satisfy the Toda lattice equation.
 
An important feature of equation \re{Id} is that, when combined with \re{Dft-spec}, it can be used to compute $D_{\mathrm{ft},\ell}(g)$ for any positive integer $\ell$. We show below (see Appendix~\ref{app:flux}) that for $\ell\ge 3$ the resulting expressions for $D_{\mathrm{ft},\ell}(g)$ are given by linear combination of harmonic polylogarithms \cite{Remiddi:1999ew,Maitre:2005uu}.

\subsection*{General case}

For a generic symbol function $\chi(x)$ (including the last two functions in \re{chis}), the determinant \re{F(g)} cannot be computed in a closed form for arbitrary $g$. However, we can derive asymptotic expansions of $D_\ell(g)$ for both small and large values of the coupling $g$, and then interpolate between them to approximate $D_\ell(g)$ at intermediate values of $g$~\cite{Belitsky:2020qir,Beccaria:2022ypy}.

At  weak coupling, the determinant \re{F(g)} has a well-defined expansion in powers of $g^2$. For symbol functions satisfying $\chi(x)\sim e^{-x}$ as $x\to\infty$, the weak coupling expansion of $D_\ell(g)$ has a finite radius of convergence, $g_\star^2=-1/16$.

The strong coupling expansion of \re{F(g)} takes a form of a transseries~\cite{Bajnok:2024epf,Bajnok:2024ymr,Bajnok:2024bqr}
\begin{align}\label{D-strong}
D_\ell(g) = \exp\bigg(-A g +  \lr{\beta\ell+\ft12 \beta^2}\log g + \mathcal F_0(g)+ \sum_{i\ge 1} g^{a_i} e^{-8\pi g x_i} \mathcal F_i(g)\bigg)\,.
\end{align}
The first term in the exponent grows linearly with $g$. Its coefficient is independent of $\ell$ and is given by the first Szeg\H{o} theorem~\cite{Szego:1915}
\begin{align}
A=2\int_0^\infty {dx\over\pi} x\partial_x \log(1-\chi(x))\,.
\end{align}
The second term in \re{D-strong} exhibits logarithmic growth with $g$ and arises from a Fisher–Hartwig singularity of the symbol function \cite{Fisher68}. It depends on both $\ell$ and the parameter $\beta$, which characterizes the behavior of the symbol near the origin, $1 - \chi(x) \sim x^{2\beta}$ for $x\to 0$. 

The sum in the exponent of \re{D-strong} involves exponentially small terms, with the exponents $a_i$ and $x_i$ uniquely determined  by the properties of the symbol function $\chi(x)$. 
The coefficient functions $\mathcal{F}_0(g)$ and $\mathcal{F}_i(g)$ (for $i \geq 1$) have asymptotic expansions in $1/g$ with factorially growing coefficients that depend on $\ell$. Consequently, these series are non-Borel summable and exhibit ambiguities, rendering individual terms in the exponent of \re{D-strong} ill-defined. However, the coefficient functions are related by nontrivial resurgence relations, which ensure the cancellation of Borel singularities across the full transseries. As a result, the expression for $D_\ell(g)$ is unambiguous and well-defined for arbitrary values of $g$. 

Requiring that the function \re{D-strong} satisfies the equation \re{Id} imposes nontrivial constraints on the coefficient functions $\mathcal{F}_0(g)$ and $\mathcal{F}_i(g)$. By solving these constraints order by order in $1/g$, we can determine the $\ell-$dependence of the coefficient functions. We have verified that the resulting expressions agree with those obtained in~\cite{Belitsky:2020qir,Beccaria:2022ypy,Bajnok:2024epf,Bajnok:2024ymr,Bajnok:2024bqr}.

\section{General solution}\label{sect3}

In this section, we construct a general solution to the equation \re{Id} satisfying the additional condition \re{bc} and discuss its properties. 

To uniquely fix the solutions to \re{Id}, we must impose an additional condition on $D_\ell(g)$. This condition arises from the small-$g$ expansion of the determinant \re{F(g)}. By changing the integration variable in \re{eq:K_nm} as $x \to g^2 x$, we find that the matrix elements $ K_{nm}(g) $ behave as  
$K_{nm}(g) = O(g^{2(n + m + \ell - 1)})$ for small $g$. This behavior allows us to expand the determinant \re{F(g)} in terms of traces of powers of the matrix $ K_{nm}(g) $, leading to the following small-$g$ expansion
\begin{align}\label{bc}
D_\ell(g) = 1 + O(g^{2(\ell + 1)}).
\end{align}
Equation \re{Id}, supplemented with the boundary condition \re{bc}, allows for a recursive determination of the function $D_{\ell+n}(g)$ for arbitrary $n \ge 1$, in terms of $D_\ell(g)$ and $D_{\ell-1}(g)$.

A general solution to \re{Id} looks as
\begin{align}\label{sol}
D_{\ell+1}(g)=2\ell D_{\ell-1}(g) \int_0^1 dx\, x^{2\ell-1} \lr{D_\ell (x g)\over D_{\ell-1}(x g)}^2.
\end{align}
In virtue of \re{ell-cond}, this relation holds for $\ell>-1-\beta$, where $\beta$ controls the behaviour of the symbol function at the origin, $\chi(x)\sim x^{2\beta}$. Subsequently applying \re{sol}, we can express $D_{\ell+n}(g)$ for arbitrary positive integer $n$ in terms of the functions $D_{\ell-1}(g)$ and $D_{\ell}(g)$. These functions plays the role of the initial conditions for the  equations \re{Id}.

The resulting expressions for $D_{\ell+n}(g)$ become rather complicated for $n\ge 2$ due to a nonlinear dependence of the right-hand side of \re{sol} on these functions. 
In spite of this we show below that these functions have remarkably simple iterative structure closely related to Dyck paths.

\subsection{Differential equations}

Computing the functions $D_{\ell+n}(g)$ we examine separately even and odd $n$ and introduce notation for their ratios 
\begin{align}\label{d-def}\notag
{}& {D_{\ell+2n-1}\over D_{\ell-1}} =\mathcal N_{2n} {d_{2n}(g)\over g^{2n(n+\ell-1)}}\,,
\\ 
{}& {D_{\ell+2n}\over D_\ell} = \mathcal N_{2n+1}{d_{2n+1}(g)\over g^{2n(n+\ell)}}\,,
\end{align}
where $n\ge 0$ and the normalization factor $\mathcal N_n$ was introduced for convenience. It is given by
\begin{align}\label{N-fact}
\mathcal{N}_n = \prod_{i=0}^{n-1} \frac{\Gamma(\ell+i)}{\Gamma(\ell)} = \frac{\ell^{n-1}}{\Gamma^{n-2}(\ell+1)} \, \frac{G(\ell+n)}{G(\ell+2)}\,,
\end{align}
where $ G(z) $ denotes the Barnes $ G $-function satisfying $G(z+1)=G(z)\Gamma(z)$. 

The functions $d_n(g)$ defined in \re{d-def} satisfy the normalization condition
\begin{align}\label{ini}
d_0=d_1=1\,.
\end{align}
Substitution of \re{d-def} into \re{Id} leads to the system of coupled differential equations for $d_n=d_n(g)$ 
\begin{align}\notag\label{eqs}
{}& d_{2n}  d'_{2n+2}-d_{2n+2}  d'_{2n}= d_{2n+1}^2 \,(\log f_+)'\,,
\\[2mm] 
{}& d_{2n-1} d'_{2n+1}-d_{2n+1} d'_{2n-1}= d_{2n}^2 \,(\log f_-)'\,,
\end{align}
where prime denotes a derivative with respect of the coupling $g$ and the notation was introduced for the functions $f_\pm=f_\pm(g)$ satisfying
\begin{align}\notag\label{f+-}
{}& (\log f_+(g))'=  2g^{2\ell-1} {D_\ell^2(g)\over D_{\ell-1}^2(g)}\,, 
 \\
{}& (\log f_-(g))'= 2g^{1-2\ell} {D_{\ell-1}^2(g)\over  D_{\ell}^2(g)} \,.
\end{align}
The  functions $f_\pm(g)$ depend on the initial conditions, $D_{\ell-1}(g)$ and $D_{\ell}(g)$, and they satisfy the relation
$
(\log f_+)'(\log f_-)'=4
$. Note that the derivatives $(\log f_\pm(g))'$ are positive definite functions for $g>0$.
  
Let us examine the equations \re{eqs} for the first few values of $n$.
For $n=0$ and $n=1$ we apply the the first and second equation in \re{eqs}, respectively, together with \re{ini} to find
\begin{align}\label{f2}\notag
{}& d_2(g)= \int_0^g d g_1 (\log f_+(g_1))' = \int_0^g d  \log f_+ (g_1) \,,
 \\
 {}& d_3(g) = \int_0^g d\log f_-(g_1)\, d_2^2(g_1)\,.
\end{align}
The lower limit of integration in these relations follows from the requirement of $d_n(g)$ to vanish for $g\to 0$. Indeed, it follows from \re{bc} that at small $g$ the function $d_n(g)$ vanishes for $n\ge 2$. 

Combining together these relations, we can represent $d_3(g)$ as an iterated integral
\begin{align}\label{f3}
d_3(g) = 2\int_0^g d\log f_-(g_1) \int_0^{g_1}d\log f_+(g_2)  \int_0^{g_2}d\log f_+(g_3) \,.
\end{align}
Next, we use the first equation in \re{eqs} for $n=1$ to find
\begin{align}\label{d4-ini}\notag
d_4(g){}&=d_2(g) \int_0^g dg_1 {d_{3}^2(g_1)\over d_2^2(g_1) }  (\log f_+(g_1))'
\\
{}&=d_2(g)   \int_0^g dg_1 {d_{3}^2(g_1)\over d_2^2(g_1) }  d_2'(g_1)\,,
\end{align}
where in the second relation we took into account the first relation in \re{f2}. We can further simplify the relation \re{d4-ini} by integrating by parts and taking into account \re{f2}
\begin{align}\label{d4-lin}
d_4(g)=-d_3^2(g) +2 d_2(g)\int_0^g dg_1\, d_2(g_1)d_3(g_1) (\log f_-(g_1))' \,.
\end{align}  
Unlike \re{d4-ini}, the right-hand side involves a multi-linear combination of the functions $d_2$ and $d_3$, each of which is given by an iterated integral \re{f2} and \re{f3}. We demonstrate in the next subsection that $d_4$ can also be expanded into a linear combination of iterated Chen integrals. 

\subsection{Iterated integrals}

Let us define the $k-$fold iterated integrals by integrating the product of the derivatives $(\log f_{\pm}(g))'$, as defined in \re{f+-}, over the simplex $0 \leq g_k \leq \cdots \leq g_1 \leq g$
\begin{align}\label{chen}
I_{\sigma_1\sigma_2\cdots \sigma_k}(g) = \int_0^g d\log f_{\sigma_1}(g_1)\int_0^{g_1} d\log f_{\sigma_2}(g_2) \dots \int_0^{g_{k-1}} d\log f_{\sigma_k}(g_k)\,.
\end{align}
These integrals are parameterized by a sequence $(\sigma_1 \sigma_2 \cdots \sigma_k)$ of signs $\sigma_i = \pm$. As implied by the definition, they satisfy a recursive relation
\begin{align} \label{chen1}
I_{\sigma_1\sigma_2\cdots \sigma_k}(g) {}&= \int_0^g d\log f_{\sigma_1}(g_1)\, I_{\sigma_2\cdots \sigma_k}(g_1) \,.
\end{align}
The positivity of the derivatives \re{f+-} implies that the functions $I_{\sigma_1\sigma_2\cdots \sigma_k}(g)$ are positive definite monotonically increasing with the coupling.

A distinguished property of the iterated integrals \re{chen} is that their product can be expanded into a linear combination of iterated integrals using the shuffle product 
\begin{align}\label{shuffle}
I_{\sigma_1\sigma_2\cdots \sigma_k}(g) I_{\sigma_{k+1}\cdots \sigma_{k+m}}(g) =\sum_{\sigma'\in (k,m)~\text{shuffles}}
I_{\sigma'_1\sigma'_2\cdots \sigma'_{k+m}}(g)\,.
\end{align}
The sum on the right-hand side runs over permutations of $(\sigma_1,\dots,\sigma_{k+m})$ such that $\sigma_1,\dots,\sigma_k$ and $\sigma_{k+1},\dots,\sigma_{k+m}$ always appear ordered. 

It is convenient to assign to the integrals \re{chen} transcendental weights $(k_+, k_-)$, where $k_+$ and $k_-$ represent the total number of positive $(+)$ and negative $(-)$ entries, respectively, in the sequence $(\sigma_1\sigma_2\cdots \sigma_k)$. 
The sum $k=k_++k_-$ gives the total weight of $I_{\sigma_1\sigma_2\cdots \sigma_k}(g)$. It follows from \re{shuffle} that  
the weight of the product of the iterated integrals is given by the sum of their individual weights.

Using the integrals \re{chen} we can rewrite \re{f2} and \re{f3} as
\begin{align}\notag\label{sol1}
{}& d_2(g)= I_+(g)\,,
\\[2mm] 
{}& d_3(g)=2  I_{-++}(g)\,.
\end{align}
Applying the relation \re{shuffle}, the product of these functions can be expanded over the basis of integrals \re{chen}, e.g.
\begin{align}\notag
{}& d_2(g) d_3(g)=2\left( 3I_{-+++} +I_{+-++}\right)\,,
\\[2mm]
{}& d_3^2(g)= 8\left(6 I_{--++++} +3 I_{-+-+++}+I_{-++-++}\right)\,.
\end{align}
Substituting these relations into \re{d4-lin} and taking into account \re{chen1} and \re{shuffle}, we find after some algebra
\begin{align}\label{d4}
d_4(g)=12 I_{+--+++}(g)+4I_{+-+-++}(g)\,.
\end{align}
According to the definition \re{chen} and \re{f+-}, the integrals entering the relations \re{sol1} and \re{d4} depend on the functions $D_{\ell-1}(g)$ and $D_\ell(g)$ in a nontrivial way. 

Substituting the relations \re{sol1} and \re{d4} into \re{d-def}, we can determine the functions $D_{\ell+n}(g)$ for $n=1,2,3$. In the next subsection, we construct a general solution to \re{eqs}. 

\subsection{Ansatz} 

By examining relations~\re{sol1} and~\re{d4}, we observe an interesting pattern: the functions $d_n$ (for $n = 2, 3, 4$) appear as linear combinations of the integrals $I_{\sigma_1\sigma_2\dots}$ with $\sigma_i = \pm$. Notably, both the total number of $\sigma$'s and the number of `$+$' and `$-$' entries within the sequence $(\sigma_1, \sigma_2, \dots)$ are correlated with the value of $n$.

This suggests to look for a general solution to \re{eqs} as a linear combination of the iterated integrals \re{chen}\,\footnote{I would like to thank Johannes Henn and members of his group at the MPI for the inspiring discussion on this point.}
\begin{align}\label{ansatz}
d_n(g)=\sum c_{\sigma_1\sigma_2\dots\sigma_k} I_{\sigma_1\sigma_2\cdots\sigma_k}(g) \,,
\end{align}
where the sum runs over sequences 
$(\sigma_1\sigma_2\dots\sigma_k)$ containing $k_+$ entries `$+$' and $k_-$ entries `$-$'.  The total number of terms in the sum \re{ansatz} is given by $(k_++k_-)!/(k_+! k_-!)$, but, as we show below, the number of contributing terms is smaller. The explicit expressions of $k_+$ and $k_-$ depend on $n$ (see \re{pq} below).

The expansion coefficients $c_{\sigma_1\sigma_2\dots\sigma_k}$ are independent of the coupling constant $g$ and parameter $\ell$, but they do depend on $n$. Substituting the ansatz~\eqref{ansatz} into \re{eqs}, we find  a system of equations for $c_{\sigma_1\sigma_2\dots\sigma_k}$. These coefficients are accompanied by bilinear combinations of the iterated integrals $I_{\sigma_1\sigma_2\dots}(g)$. 

In what follows we assume that for different sequences $(\sigma_1\sigma_2\dots)$, the corresponding integrals $I_{\sigma_1\sigma_2\dots}(g)$ are independent functions of the coupling $g$. The product of these integrals can be expanded into linear combinations of $I$'s using the shuffle identity~\re{shuffle}. By requiring both sides of Eq.~\eqref{eqs} to have the same transcendental weight, we can determine the values of $k_+$ and $k_-$ which count the number of `$+$' and `$-$' entries in the sequences  that appear  in the ansatz~\eqref{ansatz}. We find that the corresponding values of $k_+$ and $k_-$ depend on the parity of $n$
\begin{align}\notag\label{pq}
{}& n= 2 p : && \hspace*{-20mm} k_+=p^2\,, && \hspace*{-20mm} k_-= p(p-1)\,, 
\\[2mm]
{}& n=2p+1: &&  \hspace*{-20mm} k_+=p(p+1)\,, && \hspace*{-20mm}  k_-= p^2 \,. 
\end{align} 
where $p\ge 1$. In both cases, we have
\begin{align}\label{k-n}
k=k_++k_-=\frac12 n(n-1)\,.
\end{align}
We verify that for $p=1,2$ this agrees with \re{sol1} and \re{d4}. 
 
\subsection{Solutions} 

To determine the coefficients $ c_{\sigma_1\sigma_2\dots\sigma_k} $, we substitute the ansatz~\re{ansatz} into equation~\re{eqs}, apply the shuffle relation~\re{shuffle}, and match the coefficients of the functions $ I_{\sigma_1\sigma_2\dots}(g) $ on both sides of~\re{eqs}. This procedure reveals that the number of such functions exceeds the number of terms in the ansatz \re{ansatz}, leading to an overdetermined linear system for the coefficients $ c_{\sigma_1\sigma_2\dots\sigma_k} $. The excess of equations over unknowns grows rapidly with $ n $. 
Given the large number of consistency conditions imposed on the ansatz \re{ansatz}, it is a priori unlikely that generic solutions to \re{eqs} admit this form. It is therefore remarkable that the exact solutions for $d_n$ (with $n\ge 5$) that we present in this subsection are indeed captured by the ansatz \re{ansatz}.
 
Following the procedure outlined above, we used \texttt{Mathematica} to compute the expansion coefficients in~\re{ansatz} for $n = 5, 6, 7$, and to construct the corresponding functions $d_n(g)$. The resulting expression for $d_5(g)$ is
 \begin{align}\notag\label{sol2}
d_5(g)  ={}&16  
\Big(I_{-+-+-++-++}+I_{-++-+-+-++}+I_{-+-++-+-++}
\\{}&\notag 
   +3I_{-+-+-+-+++}+3I_{-+-++--+++}+3I_{-++-+--+++}+3I_{-++--++-++}
\\{}&   
   +6 I_{-+-+--++++}+9 I_{-++--+-+++}+18I_{-++---++++}\Big),
\end{align}
where the $I-$functions are defined in \re{chen}.
The corresponding expression for $d_6(g)$ is provided in Appendix~\ref{app:d6}. The result for $d_7(g)$ is too lengthy to include here and is instead available in an ancillary file.

The following comments are in order. 

We recall that the total number of terms in the sum \re{ansatz} equals $(k_++k_-)!/(k_+! k_-!)$ where $k_\pm$ are given by \re{pq} depending on the parity of $n$. Upon solving \eqref{eqs}, we observed that numerous coefficients $c_{\sigma_1\sigma_2\dots\sigma_k}$ vanished. For different values of $n$, the total number of coefficients in~\re{ansatz} and the number of non-zero coefficients are summarized below:

\begin{table}[h!]
\centering
\begin{tabular}{>{\centering\arraybackslash}p{20mm}|*{7}{>{\centering\arraybackslash}p{14mm}}}
$n$ & 2 & 3 & 4 & 5 & 6 & 7 & 8 \\
\hline
Total   & 1 & 3 & 15 & 210  & 5005  &  293930 & 30421755 \\
Non-zero   & 1 & 1 & 2 & 10 & 120 & 3276 & 197148 \\
\end{tabular}
\caption{Total and non-zero number of coefficients in~\re{ansatz} for various values of $n$.}
\label{tab:coefficients}
\end{table}

\noindent
In the next section, we show that these numbers admit a simple interpretation in terms of path counting on a square lattice.

Examining the relations \re{sol1}, \re{d4} and \re{sol2} (see also \re{sol3}) we observe that the functions $d_n$ are given by linear combinations of the iterated integrals \re{chen} in which the first two and the last two entries in the sequence $(\sigma_1,\dots,\sigma_k)$ fixed as
\begin{align}\label{ans1}\notag
{}& d_{n}(g) = \sum_{\bm\sigma} c_{+- \bm\sigma ++} I_{+- \bm\sigma ++}(g) \,,\qquad \text{for $n=$ even}
\\[2mm]
{}& d_{n}(g)= \sum_{\bm\sigma} c_{-+ \bm\sigma ++} I_{-+ \bm\sigma ++}(g) \,,\qquad \text{for $n=$ odd}
\end{align}
where the normalization factor is given by \re{N-fact} and the sum run over the sequences $\bm\sigma=(\sigma_3, \dots,\sigma_{k-2})$ containing $(k_+-3)$ entries `$+$' and $(k_--1)$ entries `$-$', see \re{pq}.

Remarkably, all expansion coefficients in \re{ansatz} and \re{ans1} are positive integers, independent of $\ell$. For arbitrary $n$, they satisfy the relation
\begin{align}\label{c-bounds}
2^{ \lfloor {(n-1)^2/ 4}\rfloor} \le c_{\sigma_1\sigma_2\dots\sigma_k} \le G(n+1)\,,
\end{align}
where  $\big\lfloor {x}\big\rfloor$ is the floor function and $G(z)$ is the Barnes $G-$function.
The maximal expansion coefficients in \re{ans1} correspond to the sequences that alternate between blocks of consecutive pluses $(+)$ and minuses $(-)$, with each block increasing in length by one at each step, e.g.
\begin{align}\label{c-largest}\notag
{}& c_{-++---++++-----++++++}\,,
\\
{}& c_{+--+++----+++++------+++++++} \,,  
\end{align}
for $n=7$ and $n=8$, respectively. The sequences corresponding to the minimal expansion coefficients have more complicated structure, e.g. 
\begin{align}\notag\label{c-smallest}
{}& c_{-++-+-+-++-+-+-+-+-++}\,,
\\[2mm]
{}& c_{+ - + - + + - + - + - + - + + - + - + - + - + - + - + +}\,,
\end{align}
for $n=7$ and $n=8$, respectively.  We describe their properties in the next section (see Figures~\ref{max-path} and \ref{min-path}). 
   
 \subsection{Sum rules}  
 
The relations \re{ans1} define the solutions to the differential equations \re{eqs}. Being combined with \re{d-def}, they allow us to compute the functions $D_{\ell+n}(g)$, defined in \re{F(g)}, for arbitrary $n\ge 1$ in terms of the initial conditions $D_{\ell-1}(g)$ and $D_\ell(g)$. 

The solutions \re{ans1} are unique specified by the set of nonzero expansion coefficients $c_{+- \bm\sigma ++}$ and $c_{-+ \bm\sigma ++}$. Most importantly, these coefficients are \textit{universal}: they depend neither on $\ell$ nor on $g$, and are independent of the choice of the initial conditions $D_{\ell-1}(g)$ and $D_\ell(g)$.  In the relations \re{ans1}, the dependence on these parameters enters solely through the the $I-$functions defined in \re{chen}.

This universality can be exploited to derive nontrivial relations for the $c-$coefficients. As demonstrated in the previous section, for specific choices of the symbol function, the function $D_\ell(g)$ admits a closed-form expression for arbitrary $\ell$ and $g$ (see \re{D-TW} and \re{D-Bessel}). Substituting these explicit expressions into \re{eqs}, we can compute the corresponding functions $d_n$ and match them to the expected form \re{ans1}. By further replacing the $I-$functions in \re{ans1} with their explicit expressions \re{chen} and \re{f+-}, we thereby obtain nontrivial sum rules for the universal expansion coefficients.
 
In particular, we show in Appendix~\ref{app:sum} that the total sum of the coefficients 
is given by
\begin{align}\label{sum-rule}
\sum c_{\sigma_1\sigma_2\dots\sigma_k}=\frac{\pi ^{n/2} \Gamma\left(\frac{1}{2}(n-1)n + 1\right) G\left(\frac{1}{2}\right)}{2^{(n-1)n/2} G\left(n + \frac{1}{2}\right)} = \{1, 2, 16, 768, 292864, 1100742656\},
\end{align}
where the numerical values on the right-hand side correspond to $n=2,3,\dots,7$.
According to  \re{ansatz}, the same sum can be recovered from the ratio
$d_n(g)$ by setting all $I-$functions to $1$. Using the explicit results \re{sol1}, \re{d4}, \re{sol2} and \re{sol3}, we have verified the identity \re{sum-rule} for these values of~$n$.    

\begin{figure}[t!]
\begin{centering}
\includegraphics[width=0.45\textwidth]{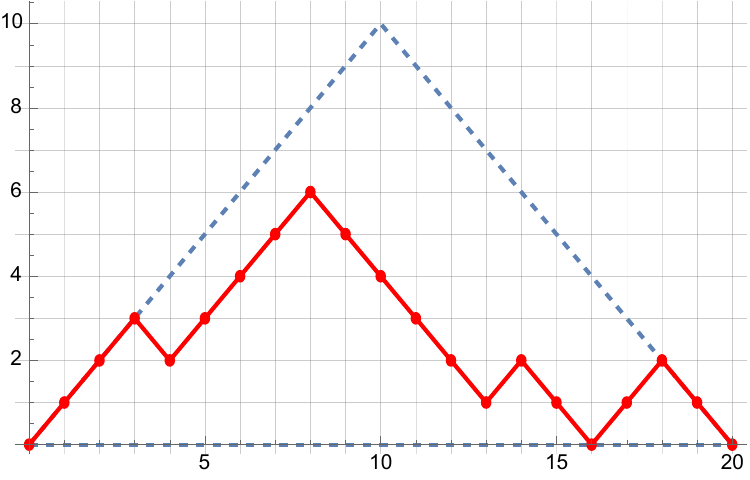}
\par\end{centering}
\caption{Example of the Dyck path that starts at $(0,0)$ and ends at $(20,0)$. The dashed line represents the boundary that the path cannot cross. The path corresponds to the sequence
$
(+++-++++-----+--++--) 
$.
}
\label{Dyck-path}
\end{figure}
  
\section{Relation to lattice paths}\label{sect:paths}  

As explained in the previous section, the functions $d_n(g)$ defined in \re{d-def} and \re{ansatz} are uniquely specified by the set of universal coefficients $c_{\sigma_1\sigma_2\dots\sigma_k}$ parameterized by the sequence of $\sigma_i=\pm$ of length $k=n(n-1)/2$. These coefficients take positive integer values and only depend on nonnegative integer $n$.

In this section, we demonstrate that the properties of the coefficients $c_{\sigma_1\sigma_2\dots\sigma_k}$ can be understood by uncovering a correspondence between the expansion \re{ansatz} and certain set of paths on a two-dimensional square lattice.\,\footnote{This interpretation was proposed by Philippe Di Francesco.}  We argue in this section that the function $d_n(g)$ can be interpreted as a partition function (or generating function in enumerating combinatorics) of  
an ensemble of lattice paths confined to a nontrivial domain. 

\begin{figure}[t!]
\begin{centering}
\includegraphics[width=0.75\textwidth]{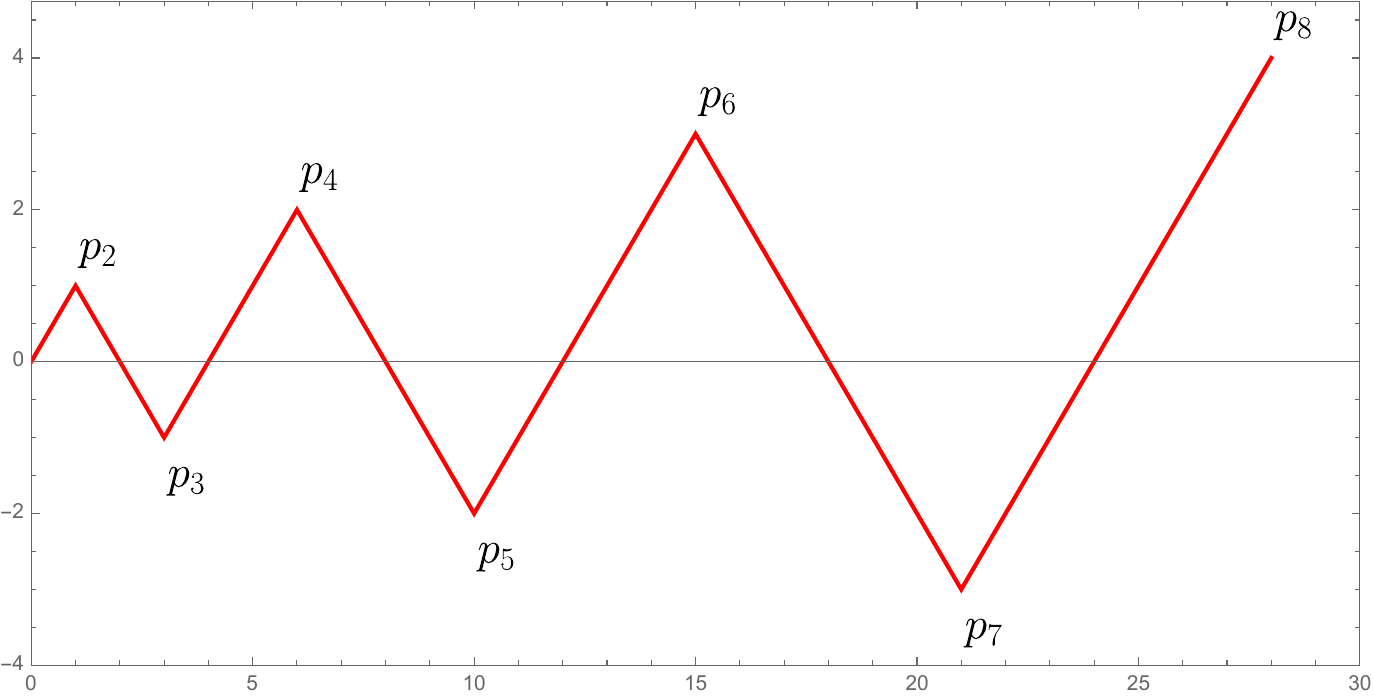} 
\par\end{centering}
\caption{The path corresponding to the maximal coefficient at \( n = 8 \). Its extrema are located at the points $p_n$ defined in \re{end-points} below. The paths for the maximal coefficients at \( n = 7, 6, 5,\dots \) start at the origin and terminate at the point $p_n$.  
}
\label{max-path}
\end{figure}

A particular example of a lattice path that is relevant for our discussion is a \emph{Dyck path}. An example of this path is shown in Figure~\ref{Dyck-path}. It starts at the origin \((0,0)\), ends on the \(x\)-axis, and never dips below it. The path consists of an equal number of up-steps (\(\nearrow\)) and down-steps (\(\searrow\)), each of length \(\sqrt{2}\). By assigning signs \(+\) and \(-\) to the up- and down-steps, respectively, we can represent a Dyck path as a sequence \((\sigma_1\sigma_2\dots\sigma_{k})\), where \(\sigma_i = \pm\) corresponds to the direction of the \(i\)th step. The total number of steps $k$ determines the horizontal coordinate of the endpoint $(k, 0)$.

To establish the connection between \re{ansatz} and lattice paths, we adopt the same convention and interpret each term in the sum \re{ansatz} as corresponding to a path on the square lattice, uniquely determined by the sequence $(\sigma_1\sigma_2\cdots\sigma_k)$. 

\subsection{Maximal and minimal coefficients}

As an example, we consider the special sequences that appear in \re{c-largest} and \re{c-smallest}. We recall that they correspond to the largest and smallest coefficients, respectively, in the sum \re{ansatz} for even and odd $n$.  

We begin with the second coefficient in \re{c-largest}, defined for even $n$. It involves the sequence of length $k = {n(n-1)}/{2}$, containing $k_+ = {n^2}/{4}$ pluses and $k_- =  {n(n-2)}/{4}$ minuses (see \re{pq}). According to the convention introduced above, we map this sequence to the path shown in Figure~\ref{max-path}. The path starts at the origin $(0,0)$ and ends at $\left( n(n-1)/2,n/2 \right)$. It consists of $k_+$ up-steps ($\nearrow$) and $k_-$ down-steps ($\searrow$). Because the sequence alternates between consecutive blocks of `$+$' and `$-$', the resulting path has a characteristic zig-zag shape composed of $(n - 1)$ segments. Each segment forms a straight run of steps, with the length of successive segments increasing by $\sqrt 2$. The longest segment, consisting entirely of up-steps, has length $(n - 1)\sqrt 2$ and is attached to the endpoint of the path.

The first coefficient in~\re{c-largest}, defined for odd~$n$, corresponds to a sequence that begins with `$-$' and ends with a block of `$+$'s of length~$(n - 1)$. Applying the sign-flipping transformation\,\footnote{Geometrically, this transformation reflects the corresponding path across the $x-$axis.}
\begin{align}\label{flip}
(\sigma_1, \sigma_2, \dots, \sigma_k) \, \stackrel{n = {\rm odd}}{\longrightarrow} \, (-\sigma_1, -\sigma_2, \dots, -\sigma_k)\,,
\end{align}
to this sequence yields a new sequence that closely resembles the one shown in Figure~\ref{max-path}. The primary difference is that the final, longest segment in Figure~\ref{max-path} must be removed. In this sense, for odd~$n$, the path corresponding to the transformed sequence~\re{flip} can be interpreted as a truncated version of the path associated with even $(n + 1)$. 

\begin{figure}[t!]
\begin{centering}
\includegraphics[width=0.65\textwidth]{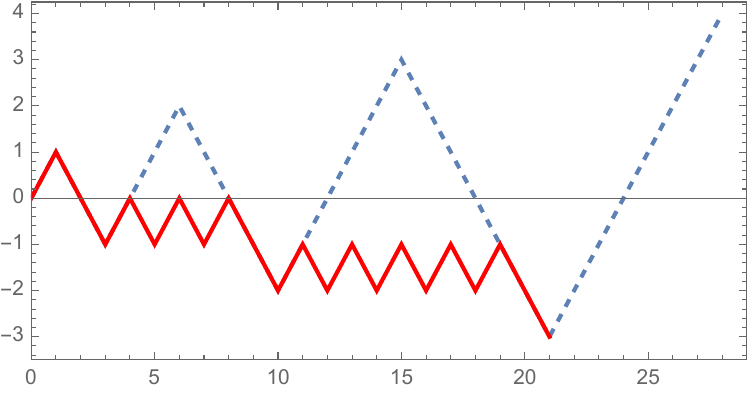}

\includegraphics[width=0.65\textwidth]{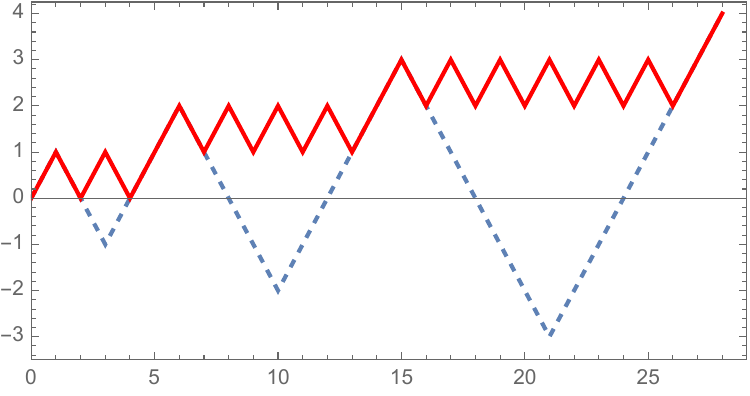}
\par\end{centering}
\caption{The paths describing the minimal coefficients at  $n=7$ (upper panel) and $n=8$ (lower panel). The dashed line depicts the path shown in Figure~\ref{max-path}.}
\label{min-path}
\end{figure} 

More generally, by successively removing the longest segments from the path in Figure~\ref{max-path}, we obtain the paths corresponding to the maximal coefficients at $(n - 1)$, $( n - 2 )$, and so on. Thus, the sign-flipping transformation \re{flip} provides a unified framework for describing the paths associated with both even and odd \( n \).  In the figures below, the transformation \re{flip} is tacitly applied for all odd \( n \).
 
In contrast to the maximal coefficients, which are universally represented by a zig-zag path for any $n$ (see Figure~\ref{max-path}), the minimal coefficients  \re{c-smallest} correspond to two distinct paths depending on the parity of $n$, as shown in Figure~\ref{min-path}. In what follows, we refer to the paths shown in Figures~\ref{max-path} and~\ref{min-path} as \textit{maximal} and \textit{minimal} weighted paths, respectively.

\subsection{Selection rule}

We are now ready to describe the paths associated with the remaining coefficients in \re{ansatz} and \re{ans1}. For a fixed value of $n$, these coefficients correspond to sequences $(\sigma_1 \sigma_2 \dots \sigma_k)$ of length $k = n(n-1)/2$, composed of $k_+$ pluses and $k_-$ minuses, where $k_\pm$ are given by \re{pq}. 
Following the conventions outlined above, each such sequence can be mapped to a specific path on the square lattice. For all coefficients appearing in~\re{ansatz}, the corresponding paths begin at the origin $(0,0)$ and terminate at the point
\begin{align}\label{end-points}
p_n=\left( \frac{n(n-1)}{2}, (-1)^n\left\lfloor\frac{n}{2}\right\rfloor \right)
\end{align}
regardless of the order in which the `$+$' and `$-$' signs appear in the sequence. Here the factor of $(-1)^n$ arises due to the sign-flipping transformation \re{flip}. Note that for $n=1$ the end-point $p_1$ coincides with the origin.

It is straightforward to verify that for $n = 2, 3, 4, \dots$, the endpoints \re{end-points} lie at the tips of the zig-zag pattern shown in Figure~\ref{max-path}. Moreover, it follows from~\re{ans1} that each path begins with an up-step followed by a down-step, and ends with a sequence of two up-steps if $n$ is even, or two down-steps if $n$ is odd. It is easy to see that the paths illustrated in Figures~\ref{max-path} and~\ref{min-path} satisfy this condition.

\begin{figure}
\begin{centering}
\includegraphics[width=0.7\textwidth]{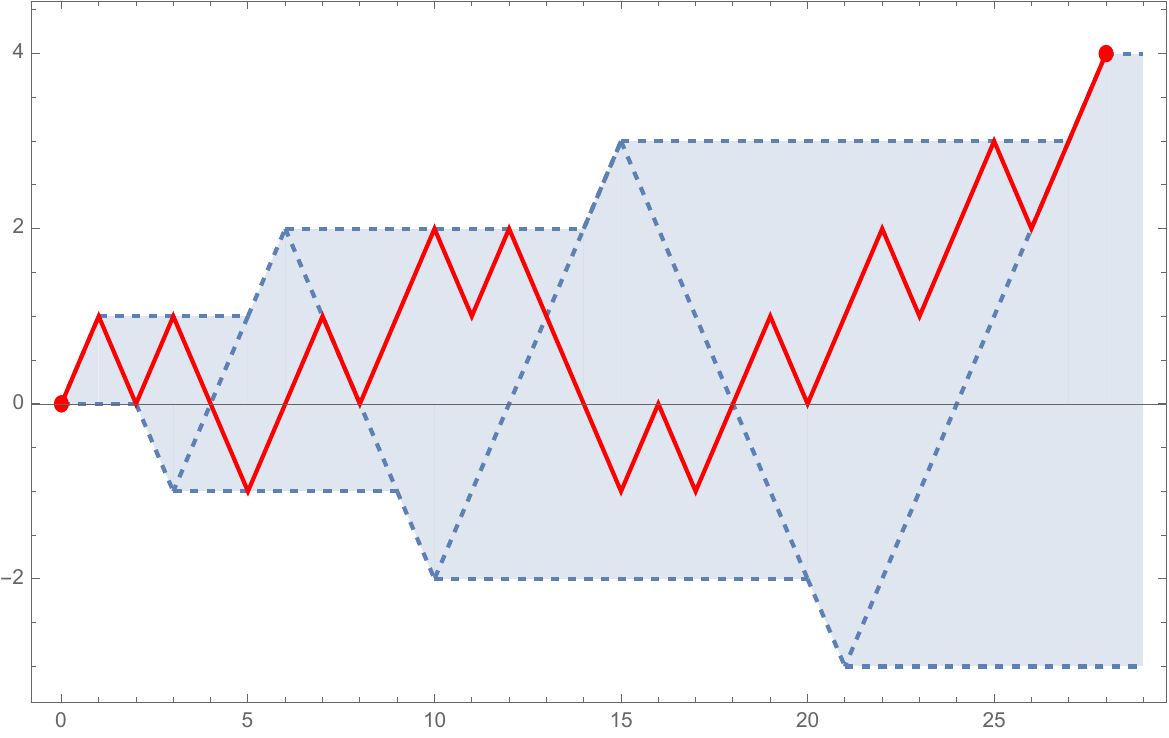}
\par\end{centering}
\caption{Envelope of admissible paths contributing to~\re{ansatz}. The red line depicts an example of a (irreducible) path for $n=8$, red dots indicate the start and end points. The allowed region is constructed by joining individual wedges along their shared edges, indicated by dashed lines.}
\label{area}
\end{figure}   

As observed earlier (see Table~\ref{tab:coefficients}), the number of nonzero coefficients in~\re{ansatz} is significantly smaller than the total number of possible coefficients. This sparsity admits a natural interpretation in terms of the corresponding lattice paths, and can be encapsulated by the following selection rule:
\begin{align} \label{rule}
 \begin{split}
\text{ 
{\parbox[c]{0.75 \textwidth}{\textit{All admissible paths must terminate at the point \re{end-points} and remain entirely within the envelope defined by the shaded region in Figure~\ref{area}. Paths that violate this rule do not contribute.
}}}}
 \end{split}
\end{align}
 
\noindent  
By counting the total number of paths that satisfy this rule, we recover the entries listed in Table~\ref{tab:coefficients}.

We observe that the minimal weighted paths shown in Figure \ref{min-path} are unique in that they follow either the lower or upper boundary of the allowed region, depending on the parity of $n$. In the similar manner, the maximal weighted path depicted in Figure \ref{max-path} (and shown in dashed lines in Figure \ref{area}) naturally cuts the allowed region into a collection of `wedges'. The entire region is formed by sewing together individual wedges along their common edges. Similarly, the allowed path is constructed by gluing together segments that belong to different wedges. 

The path segments within each individual wedge begin and end on its edges and share the same properties as the Dyck path shown in Figure~\ref{Dyck-path} -- they strictly remain within the wedge's limits.\,\footnote{To see this correspondence, some wedges must be reflected across the horizontal axis.} For this reason, we refer to them as \textit{generalized Dyck paths}.
For given $n$, the area shown in Figure \ref{area} comprises $(n-2)$ wedges. Consequently, the allowed paths are formed by concatenating $(n-2)$ generalized Dyck paths. It is worth mentioning that for certain paths, such as the maximally weighted paths in Figure~\ref{max-path}, some path segments coincide with conventional Dyck paths.

\begin{figure}
\begin{centering}
\includegraphics[width=0.8\textwidth]{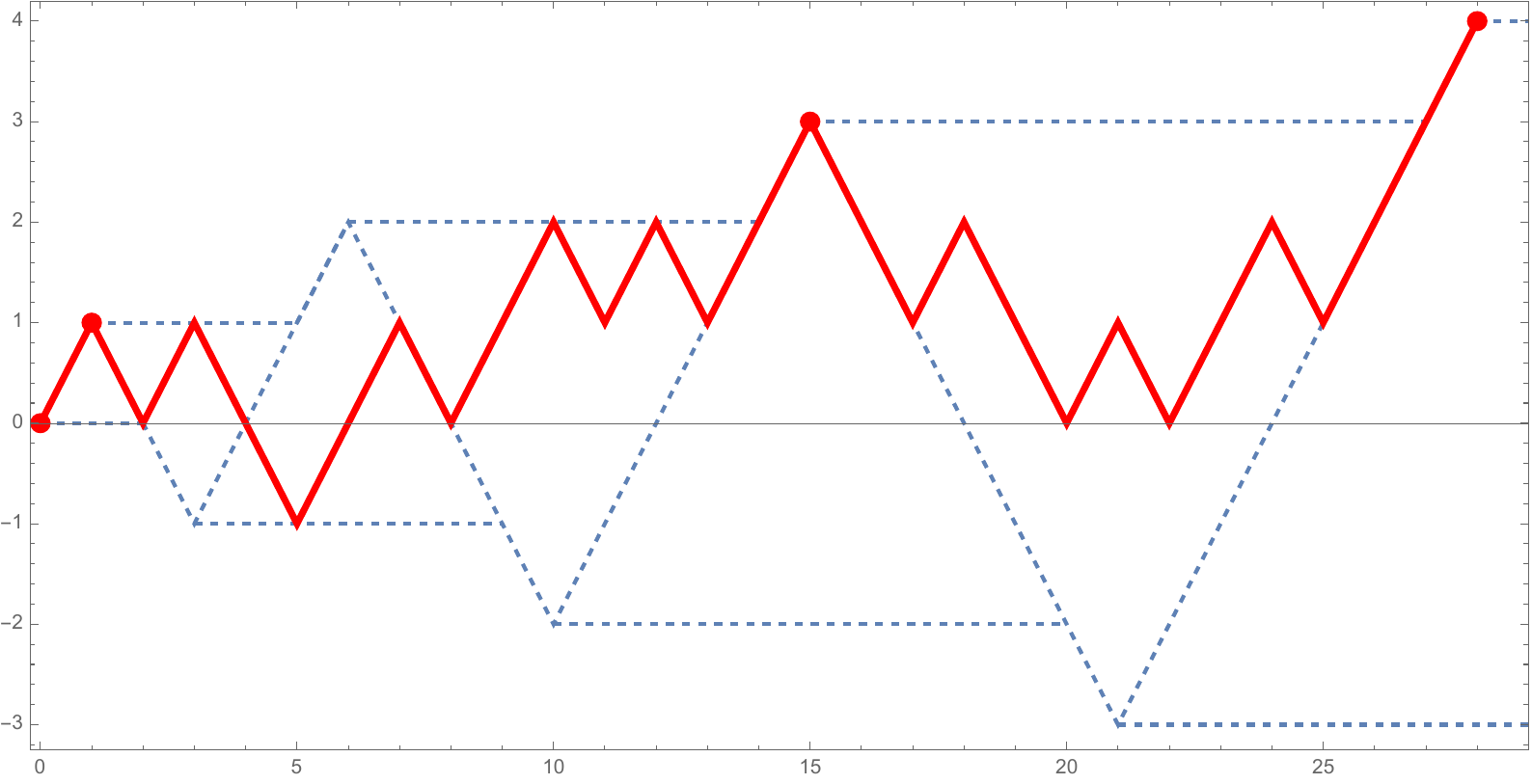}
\par\end{centering}
\caption{Example of a reducible path. Red dots mark the end-points $p_m$ defined in \re{end-points}. The path begins at $p_1$ and passes through the points $p_2$, $p_6$ and $p_8$.}
\label{red-path}
\end{figure} 

For given $n$, the allowed paths start at the origin $p_1=(0,0)$ and terminate at the end-point $p_n$ defined in \re{end-points}. If the path passes through the intermediate point $p_m$ for $m<n$, it naturally splits into the allowed path connecting the points $p_1$ and $p_m$ and the subpath connecting the points $p_m$ and $p_n$. This leads us to classify all allowed paths as either {\em irreducible} or {\em reducible}.  Irreducible paths, like the one in Figure~\ref{area}, do not pass through the intermediate points $p_m$ for $3\le m\le n-1$.\,\footnote{For $n>2$, every allowed path must pass through the point $p_2$; this explains the lower bound on $m$.} Conversely, the path in Figure~\ref{red-path} 
is reducible. 
The definition of the reducible paths is given below in \re{factorization}.
 
As we show in the next subsection, the irreducible paths play a distinguished role in our analysis. 
           
\subsection{Weights}         

We verified that the solutions of \re{eqs} for $d_n$ with $n=2,3,\dots,7$ all obey the selection rule \re{rule}. For the reader’s convenience, the explicit expressions for $d_n$ are collected in the attached {\tt Mathematica} file.  Extending these solutions to higher values of $n$ becomes increasingly challenging due to the factorial growth in the number of terms in $d_n(g)$. 
The selection rule~\re{rule} provides a powerful and efficient means to construct the solution \re{ansatz} for arbitrary $n$, circumventing the need for explicit calculations.

Each allowed path contributes to the ansatz \re{ansatz} with a positive integer coefficient $c_{\sigma_1\sigma_2\dots\sigma_k}$ that satisfies relation \re{c-bounds}. For convenience, we normalize all coefficients by the maximal one, introducing the notation 
\begin{align}\label{w}
w_{\sigma_1\sigma_2\dots\sigma_k} = {c_{\sigma_1\sigma_2\dots\sigma_k}\over G(n+1)}\,,
\end{align}
where $k=n(n-1)/2$ and $\sigma_i=\pm$.
The advantage of using these normalized coefficients is that they are positive rational values, satisfying the relation
\begin{align}\label{limits}
w_{\rm min}(n) \le w_{\sigma_1\sigma_2\dots\sigma_k} \le 1\,,
\end{align}
where $w_{\rm min}(n)=2^{ \lfloor {(n-1)^2/ 4}\rfloor}/G(n+1)$. The sum of all weights $\sum w_{\sigma_1\sigma_2\dots\sigma_k}$ can be found from \re{sum-rule}.

The coefficients \re{w} can be interpreted as statistical weights for individual paths within the partition function 
\begin{align}\label{Z}
Z_n(g)= \sum w_{\sigma_1\sigma_2\dots\sigma_k} I_{\sigma_1\sigma_2\dots\sigma_k}(g)\,,
\end{align}
where the sum runs over all paths satisfying \re{rule}.
The solution \re{ansatz} is related to this partition function as 
\begin{align}\label{d-Z}
d_n(g)=G(n+1) Z_n(g)\,.
\end{align} 
The partition function \re{Z} can be viewed as a generalization of the Dyck path partition function, which plays an important role in statistical mechanics and enumerative combinatorics \cite{Stanley_2023}. The latter is defined by summing over Dyck paths, with each path weighted according to its characteristics, such as area, height, or the number of peaks. In close analogy, the partition function \re{Z} is defined as a sum over the paths depicted in Figure~\ref{area}. Each path contributes with a weight given by the product of the coefficients \re{w} and the iterated integrals introduced in \re{chen}. These integrals depend on the path in a nontrivial way and, for general functions $f_\pm(g)$ do not admit a simple geometric interpretation. However, for a specific choice of $f_\pm(g)$ described in Appendix~\ref{app:sum}, the integrals become closely related to the height profile of the path.         

We emphasize that the partition function \re{Z} satisfies the system of differential equations derived from \re{eqs} and \re{d-Z}. In principle, by substituting \re{Z} into these equations, one could perform inverse engineering to explicitly determine the weights $w_{\sigma_1\sigma_2\dots\sigma_k}$. However, rather than pursuing this route, we analyze the explicit solutions for 
$d_n$ (with $n=2,\dots,7$) to infer the properties of the weights.~\footnote{Explicit expressions for the weights of the various paths contributing to \re{Z} are included in the {\tt Mathematica} notebook attached to this submission.}
 
\begin{figure}
\begin{centering}
\qqqquad 
\includegraphics[width=0.35\textwidth]{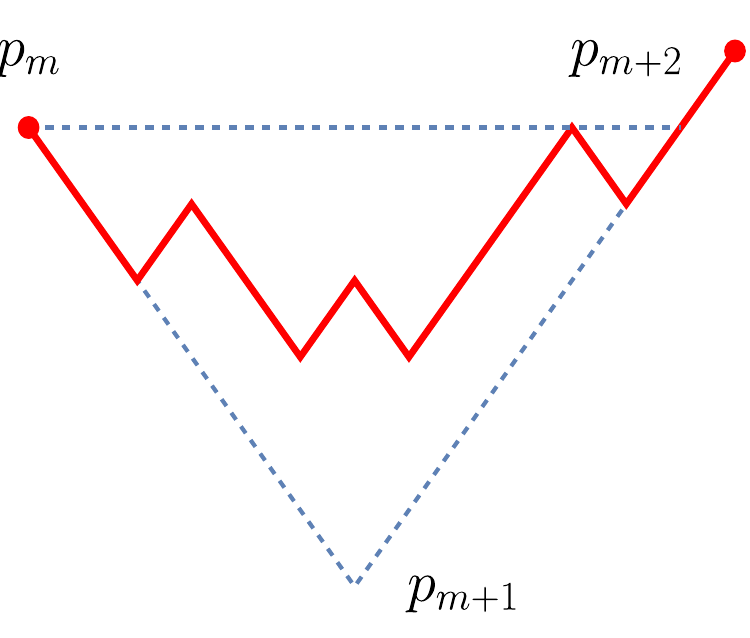} \qqquad 
\includegraphics[width=0.45\textwidth]{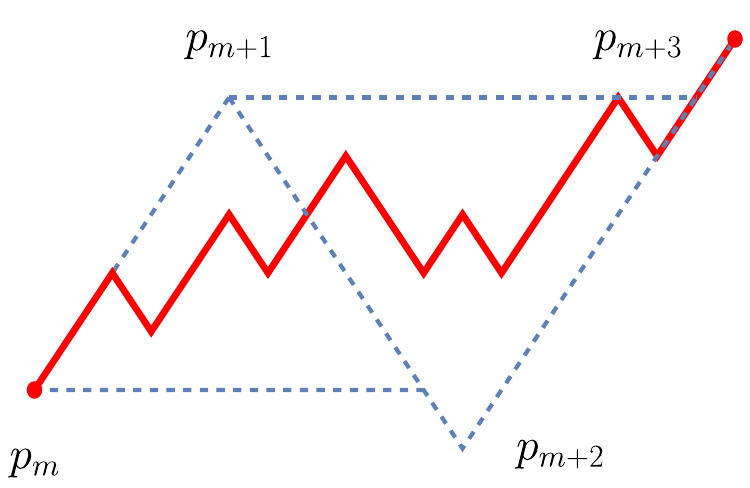} 
\par\end{centering}
\caption{Examples of the subpaths $\bm{\sigma}_{m \to m+2}$ and $ \bm{\sigma}_{m \to m+3}$ introduced in \re{factorization}. As in previous figures, red dots mark the end-points. Horizontal dashed lines indicate the boundary of the allowed region.}
\label{subpaths}
\end{figure}

Let us consider a subset of reducible paths that originate at the origin, terminate at the point \(p_n\) (as defined in \eqref{end-points}), and pass through an intermediate point \(p_m\) with $3\le m \le n-1$. Each such path can be decomposed into two segments: one from the origin to 
$p_m$, and another from $p_m$ to $p_n$. Accordingly, the associated sequence 
$\bm{\sigma}_n=(\sigma_1\sigma_2\dots\sigma_{k})$ can be represented as
\begin{align}\label{factorization}
\bm{\sigma}_n = (\bm{\sigma}_m, \bm{\sigma}_{m \to n})\,,
\end{align}
where \(\bm{\sigma}_{m \to n}\) encodes the subpath from \(p_m\) to \(p_n\). Note that $\bm{\sigma}_m=\bm{\sigma}_{1 \to m}$. Examples of the subpaths corresponding to $\bm{\sigma}_{m \to n}$ are shown in Figure~\ref{subpaths}.

The properties of the weights can be formulated as follows:
\begin{description}
\item[Property 1:]
The weights of reducible paths satisfying the condition \re{factorization} factorize into the product of the weights of their constituent subpaths
\begin{align}\label{w-factors}
w_{\bm{\sigma}_n} = w_{\bm{\sigma}_m} w_{m \to n}\,.
\end{align}

\item[Property 2:]
If the path passes through few intermediate points $p_{m_1}, p_{m_2},\dots, p_{m_r}$, the relation \re{w-factors} can be continued recursively 
\begin{align}\label{w-prod-w}
w_{\bm{\sigma}_n} = w_{1\to m_1} w_{m_1\to m_2} \dots w_{m_{r}\to n}\,,
\end{align}
where $w_{1\to m_1}\equiv w_{\bm{\sigma}_{m_1}}$. Note that the weights $w_{m \to m'}$ are \emph{local} -- they depend only on the path segment between the points \(p_m\) and \(p_{m'}\), and are independent of the rest of the path's configuration.   

\item[Property 3:]
In the special case $m=n-1$, the weight $w_{(n-1) \to n}$ corresponds to the segment connecting the two adjacent end-points
$p_{n-1}$ and $p_n$. It is given by
\begin{align}\label{w=1}
w_{(n-1) \to n} = 1\,.
\end{align}

\end{description}  
In virtue of \re{w-prod-w}, the problem of computing the weights \(w_{\bm{\sigma}_n}\) reduces to determining the elementary weights \(w_{m_1 \to m_2}\), which correspond to the allowed paths that start at the point \(p_{m_1}\) and terminate at \(p_{m_2}\).
In the next subsection, we present an explicit solution for the case \(w_{m_1 \to m_2}\) with \(m_2 = m_1 + 2\).  
  
We present below several examples illustrating the application of relations \re{w-factors} -- \re{w=1}.

Let us consider the path that passes through the intermediate point $p_{n-1}$ and terminates at $p_n$. Applying \re{w-factors} and \re{w=1} we find its weight as
\begin{align}
w_{\bm{\sigma}_n} = w_{\bm{\sigma}_{n-1}} w_{n-1 \to n} = w_{\bm{\sigma}_{n-1}} \,.
\end{align}
If the path corresponding to the sequence $\bm \sigma_{n-1}$ passes through the point  $p_{n-2}$, then the relation continues recursively. This situation arises, for instance, in the case of the maximally weighted path shown in Figure~\ref{max-path}, where $w_{\rm max}(n)=w_{\bm \sigma_n}$ satisfies
\begin{align}
w_{\rm max}(n)=w_{\rm max}(n-1)=\dots=w_{\rm max}(2)=1\,.
\end{align}
For the minimal weighted path shown in Figure~\ref{min-path} the relation \re{w-prod-w} takes different form depending on the parity of $n$,
\begin{align}\notag\label{w-min-fact}
{}& w_{\rm min}(n) = w_{1\to 2}\,w_{2\to 4} \, w_{4\to 6}\, \dots w_{n-2 \to n}\,, 
\\[2mm]
{}& w_{\rm min}(n) = w_{1\to 2} \,w_{2\to 3}w_{3\to 5} \, w_{5\to 7} \, \dots w_{n-2 \to n}\,, 
\end{align}
for even and odd $n$, respectively. Both relations involve two types of weights $w_{m\to m+1}$ and $w_{m\to m+2}$. The former are given by \re{w=1} while the latter are defined below in \re{w-omega}.
  
For the path shown in Figure~\ref{red-path}, the relation \re{w-prod-w} gives
\begin{align}
w_{\bm{\sigma}_8} = w_{1\to 2} \, w_{2\to 6} \, w_{6\to 8}\,.
\end{align}
It involves the same types of weights as \re{w-min-fact} and, in addition, a new type $w_{m\to m+4}$.

\subsection{Weights of the single wedge}

Let us consider the subpath $\bm{\sigma}_{m \to m+2}$, which starts at $p_m$ and ends at $p_{m+2}$. As discussed above, this subpath lies entirely within the bounding wedge region.

\begin{figure}
\begin{centering}
\includegraphics[width=0.45\textwidth]{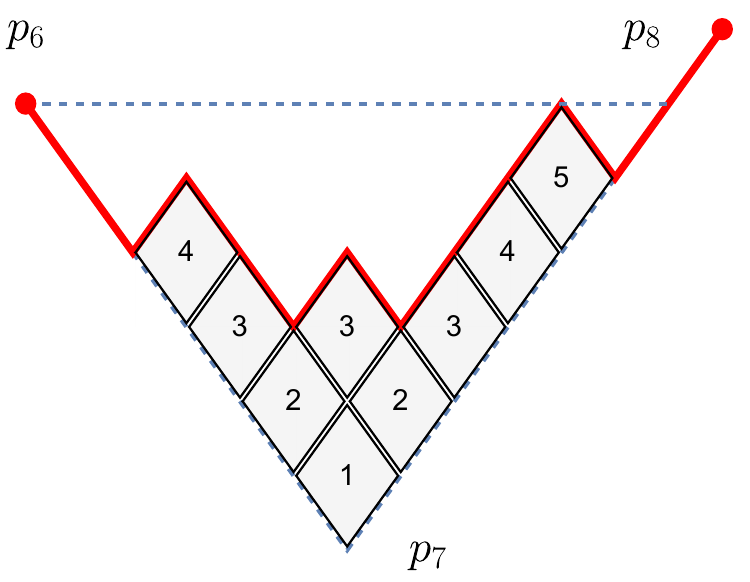} 
\par\end{centering}
\caption{Definition of the weight $w_{6\to 8}$ of the path connecting points $p_6$ and $p_8$ in terms of boxes filling the area beneath the path within the bounding wedge. Each box is labeled by an integer representing its height. This definition extends naturally to wedges reflected across the horizontal axis.}
\label{fig:boxes}
\end{figure} 

To define the corresponding weight $w_{m\to m+2}$, we adopt the method developed in \cite{DiFrancesco:1996zq} for studying meander statistics. Namely, we fill the area between the path and the boundary of the wedge with unit boxes and assign to them  height $h$ with respect to the intermediate point $p_{m+1}$ as shown in Figure~\ref{fig:boxes} for $m=6$. Since the boxes have to lie within the boundary of the allowed region, the maximal height is related to the distance between the points $p_m$ and $p_{m+1}$
\begin{align}
1\le h \le m-1\,.
\end{align}
The configuration of boxes can be described by the number of boxes \(b_i\) at each height level \(i = 1, 2, \dots, m-1\). These numbers satisfy the constraints
\begin{align}
0 \leq b_i \leq i\,, \qqqquad b_1 \leq b_2 \leq \dots \leq b_{m-1}\,.
\end{align}
For empty set of boxes, when all \(b_i = 0\), we recover the subpath corresponding to the maximally weighted path shown in Figure~\ref{max-path}. In the fully packed case, for \(b_i = i\) for all \(i\), the configuration corresponds to the subpath of the minimally weighted path shown in Figure~\ref{min-path}.
 
\begin{description}

\item[Property 4:]
Let us assign to each box at height $h$ the weight
\begin{align}
\omega_h ={m-h\over m-h+2}\,.
\end{align}
The weight $w_{m\to m+2}$ associated with the subpath $\bm{\sigma}_{m \to m+2}$ is given by the product of weights of individual boxes
\begin{align}\label{w-omega}
w_{m\to m+2} = \omega_1^{b_1} \omega_2^{b_2} \dots \omega_{m-1}^{b_{m-1}} = \prod_{i=1}^{m-1} \lr{m-i\over m-i+2}^{b_i}\,.
\end{align}

\end{description}
Applying this property in the two extreme cases of empty and fully packed box configurations mentioned above we have
\begin{align}\label{fully}
w_{m\to m+2} (b_i=0) = 1\,,\qqqquad w_{m\to m+2} (b_i=i) = {2^m\over m! (m+1)!}\,.
\end{align} 
For the empty box configuration, the path $\bm \sigma_{m\to m+2}$ splits into two segments $(\bm \sigma_{m\to m+1},\bm \sigma_{m+1\to m+2})$. Therefore, applying \re{w-prod-w} and \re{w=1} we find that $w_{m\to m+2} (b_i=0) = 1$, in agreement with \re{fully}. 

We can check the second relation in \re{fully} by computing the weights of the paths shown in Figure~\ref{min-path}. Substituting this relation into \re{w-min-fact} we obtain the expression for $w_{\rm min}(n)$ which agrees with the analogous expression in \re{limits}.

For the path shown in Figure~\ref{fig:boxes} we have $m=6$ and the number of boxes is $b_1=1$, $b_2=2$, $b_3=3$, $b_4=2$ and $b_5=1$. Applying \re{w-omega} we determine the weight associated with this path as
$w_{6\to 8}=1/175$.
 
 \subsection{Higher weights}
 
\begin{figure}
\begin{centering}
\includegraphics[width=0.45\textwidth]{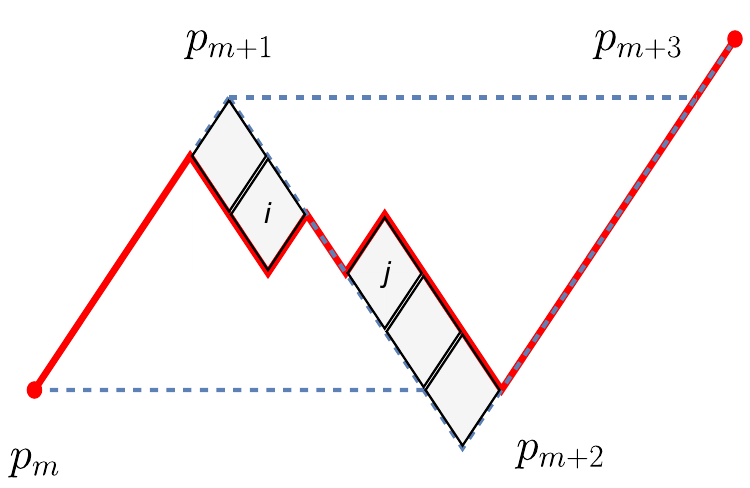} 
\par\end{centering}
\caption{Configuration of boxes contributing to the weight \( w_{m \to m+3} \). 
The number of boxes in the left and right wedges cannot exceed \( m-1 \) and \( m \), respectively, 
and their total number is further constrained by \( i + j \le m+1 \).
}
\label{int-boxes}
\end{figure}

We conclude this section with some remarks on the weights \( w_{m \to m+3} \), which correspond to the paths depicted in the right panel of Figure~\ref{subpaths}. These paths lie within the region formed by gluing together two adjacent wedges along their common edge. Similar to the case of \( w_{m \to m+2} \), the paths contributing to \( w_{m \to m+3} \) can be parameterized by two sets of boxes, located between the path and the horizontal boundary within the left and right wedges in Figure~\ref{subpaths} (see also Figure~\ref{int-boxes}).

For each set of boxes, we can apply relation~\re{w-omega} to compute the corresponding weights \( w_{m \to m+2} \) and \( w_{m+1 \to m+3} \). This naturally raises the question: how does \( w_{m \to m+4} \) compare to the product of these two weights?
By examining the values extracted from the exact solutions for \( d_n \) with \( n \leq 7 \), we find that
\begin{align}\label{w-diff}
\Delta w_{m \to m+3} \equiv w_{m \to m+3} -w_{m \to m+2} \, w_{m+1 \to m+3} \geq 0\,.
\end{align}
The difference $\Delta w_{m \to m+3}$ can be interpreted as arising from a nontrivial correlation between the two sets of boxes across the adjacent wedges.

As an illustrative example, consider the path depicted in Figure~\ref{int-boxes}. By filling the area beneath this path within the two wedges, we obtain two stacks of boxes located on opposite sides of the line connecting the points \( p_{m+1} \) and \( p_{m+2} \). Denoting the numbers of boxes in these stacks by \( i \) and \( j \), respectively, we find from \re{w-omega} that the weights associated with these boxes are
\begin{align}\notag
{}&w_{m\to m+2} = \prod_{i'=1}^{i} \lr{m-i'\over m-i'+2} =\frac{(m-i) (m-i+1)}{m(m+1)}\,,
\\
{}&w_{m+1\to m+3} = \prod_{i'=1}^{j} \lr{m+1-i'\over m-i'+3} =\frac{(m+1-j) (m-j+2)}{(m+1) (m+2)}\,,
\end{align}
where the non-negative integers \( i \) and \( j \) satisfy $0 \le i \le m-1$ and $0 \le j \le m$.

In the absence of correlations between the boxes, the weight of the path would simply be given by the product of the individual weights \( w_{m\to m+2} \, w_{m+1\to m+3} \). 
However, by examining the weight of the path shown in Figure~\ref{int-boxes} within the solution~\re{ansatz} for \( n \le 7 \), 
we find that it deviates from this naive product. 
It turns out that the difference function~\re{w-diff} for this path takes the remarkably simple form
\begin{align}\label{w-simp}
\Delta w_{m \to m+3} = \frac{i(i+1) j(j-1)}{\bigl(m(m+1)\bigr)^2}\,,
\end{align}
where the non-negative integers \( i \) and \( j \) satisfy the constraint \( i + j \le m+1 \). Consequently, the weight of the path shown in Figure~\ref{int-boxes} is given by 
\begin{align}
w_{m \to m+3} =  \frac{i(i+1) j(j-1)}{\bigl(m(m+1)\bigr)^2}+\frac{(m-i) (m+1-i)}{m(m+1)}\frac{(m+1-j) (m+2-j)}{(m+1) (m+2)}
\end{align}

The simplicity of \re{w-simp} suggests that, by analogy with \re{w-omega}, it should be possible to derive an explicit expression for \( w_{m \to m+3} \) for a generic path and, more generally, for the weights \( w_{m \to m'} \). 

A closely related problem of computing the weights of lattice paths has been extensively studied in the context of meander statistics. It was shown in \cite{DiFrancesco:1996zq} that the intricate combinatorics of meanders can be systematically encoded and evaluated using the Temperley-Lieb algebra. It would be interesting to explore whether a similar algebraic framework could be adapted to address the present problem.

\section{Scaling behaviour}

The relation \re{ansatz} is valid for arbitrary values of the coupling $g$ and for any positive integer $n$. In this section, we study the asymptotic behavior of the partition function \re{ansatz} in various regimes of $g$ and $n$.

The coupling constant governs the relative weights $I_{\sigma_1\sigma_2\cdots\sigma_k}(g)$ assigned to individual paths in \re{ansatz}. According to 
\re{chen}, these iterated integrals are built from the functions $(\log f_\pm (g))'$ defined in \re{f+-}. 
At weak coupling, we apply \re{bc} to find
\begin{align} 
(\log f_\pm (g))' = 2g^{\pm(2\ell-1)}\lr{1+O(g^{2\ell})} \,.
\end{align}
At strong coupling, we instead use \re{D-strong} to obtain  
\begin{align} 
(\log f_\pm(g))' = 2g^{\pm(2 \ell+2\beta -1)}e^{\pm B_\ell + O(1/g)}  \,, 
\end{align}
where $B_\ell$ is a constant independent on $g$. Neglecting the subleading corrections, we find that in both regimes the functions $f_\pm(g)$ have the same behaviour
\begin{align}\label{df}
 d\log f_\pm \sim  d\big(g^{h_\pm}\big)\,, 
\end{align}
where $h_+=-h_-$ takes different values of weak and strong coupling. 

Substituting equation~\re{df} into~\re{chen}, we can explicitly compute the functions \( I_{\sigma_1 \sigma_2 \dots \sigma_k} \) for an arbitrary sequence of signs \( \sigma_i \). In Appendix~\ref{app:sum}, we show that these functions admit a simple interpretation in terms of the height function associated with the lattice paths corresponding to the sequence \( (\sigma_1 \sigma_2 \dots \sigma_k) \). Consequently, in both the weak and strong coupling regimes, the function~\re{ansatz} can be interpreted as the partition function of an ensemble of lattice paths shown in Figure~\ref{area} weighted by their height.

At strong coupling, the function \re{ansatz} exhibits a nontrivial behaviour in the double scaling limit where both $g$ and $n$ tend to infinity with their ratio held fixed \cite{Bargheer:2019exp,Belitsky:2020qir}
\begin{align}\label{dsl}
n \to \infty\,, \qqqquad g \to \infty\,, \qqqquad \bar{n} = \frac{n}{2g} = \text{fixed}\,.
\end{align}
The length of the paths shown in Figure~\ref{area} is proportional to $k = n(n-1)/2$, and thus diverges in this regime. Consequently, we expect the partition function \re{ansatz} to develop a non-trivial scaling behavior.

Indeed, it was shown in~\cite{Belitsky:2020qir} that in the double scaling limit~\re{dsl}, the determinant \re{F(g)} admits the asymptotic expansion
\begin{align}\label{D-dsl}
D_{\ell+n}(g) = \exp\left(g F_0(\bar{n}) +F_1(\bar{n}) + O(g^{-1})\right)\,,
\end{align}
where the coefficient functions $F_0(\bar{n})$ and $F_1(\bar{n})$ depend on the scaling parameter $\bar{n}$ defined in~\re{dsl}. The leading-order term is given by
\begin{align}
F_0(\bar{n}) = \frac{2}{\pi} \int_{\bar{n}}^\infty \frac{dx}{x}\, \log(1 - \chi(x)) \sqrt{x^2 - \bar{n}^2}\,.
\end{align}
Substituting the expression \re{D-dsl} into the definition \re{d-def}, we can find that the asymptotic behaviour of the function $d_n(g)$ in the limit \re{dsl}.

The function \re{D-dsl} has to satisfy the equation \re{Id}, evaluated at the shifted index $\ell \to \ell + n$. It is straightforward to verify that the leading term in~\re{D-dsl} satisfies this equation identically for an arbitrary function $F_0(\bar{n})$. By examining the subleading corrections, we obtain the following constraint for the next-to-leading function $f_1(\bar{n})$ 
\begin{align}
F_1'(\bar{n}) = \frac{1}{8} F_0''(\bar{n}) \left(4\ell - \bar{n} F_0''(\bar{n})\right)\,.
\end{align}
In a similar manner, higher-order corrections to the expansion~\re{D-dsl} can be determined systematically by applying equation~\re{Id} order by order in $1/g$.

Substituting the relation \re{D-dsl} into \re{d-def} reveals a nontrivial scaling behavior of the lattice partition function \re{d-Z} in the double-scaling limit \re{dsl}. In this regime, the partition function receives the contribution of infinitely long lattice paths of the same length proportional to $n$, as illustrated in Figure~\ref{area}. It would be interesting to derive the scaling behavior \re{D-dsl} directly from the definition of the partition function \re{Z}.
  
\section*{Acknowledgments}
 
I would like to thank Philippe Di Francesco, Sasha Goncharov, Emmanuel Guitter, and Johannes Henn for many illuminating discussions. This work  was supported by the French National Agency for Research grant ``Observables'' (ANR-24-CE31-7996).  

\appendix

\section{Proof of the equation \re{Id}}\label{App:A}

In this appendix, we present a derivation of the equation \re{Id}. The function $D_\ell(g)$ entering this equation is given by the Fredholm determinant \re{F(g)} of the semi-infinite matrix \re{eq:K_nm}. 

It is convenient to interpret the matrix \re{eq:K_nm} as representing a certain integral operator on a linear space spanned by the functions \re{psi}.  
To this end, we introduce two operators
\begin{align}\notag\label{kernels}
{}& \bm K_\ell  f(x) = \int_0^\infty dy\, K_\ell(x,y) f(y)\,,
\\
{}& \bm \chi f(x) =\chi\Big({\sqrt x\over 2g}\Big) f(x)\,,
\end{align}
where $f(x)$ is a test function and the second relation involves the symbol function.  In the first relation, the kernel $K_\ell(x,y)$ is given by\,\footnote{In what follows, we use boldface  notation for the operators to distinguish them from their integral kernels.}
\begin{align}\notag\label{K-sum}
K_\ell(x,y) {}&= \sum_{n\ge 1} \psi_n(x)\psi_n(y) 
\\  
{}&= \frac{\sqrt{y} J_{\ell}\left(\sqrt{x}\right)
   J_{\ell-1}\left(\sqrt{y}\right)-\sqrt{x} J_{\ell-1}\left(\sqrt{x}\right)
   J_{\ell}\left(\sqrt{y}\right)}{2 (x-y)}\,, 
\end{align}
where the functions $\psi_n(x)$ are defined in \re{psi} and the second relation follows from the Christoffel-Darboux formula for these functions. According to the definition \re{kernels}, the operator $\bm K_\ell$ is independent on the coupling $g$, whereas $\bm \chi f(x)$ is independent on the parameter $\ell$.

It follows from the orthogonality condition of the functions \re{psi} that $\bm K_\ell\, \psi_n(x)=\psi_n(x)$.
We can employ quantum-mechanical notations to represent the entries of the matrix \re{eq:K_nm} as matrix elements
\begin{align}\label{trun}
K_{nm}(g)=\vev{\psi_n| \bm \chi |\psi_m}=\vev{\psi_n|{\bm K}_\ell\, \bm \chi |\psi_m} \,.
\end{align}
According to \re{kernels}, the product of the operators ${\bm K}_\ell\,\bm \chi$ has a kernel $K_\ell(x,y) \chi(\sqrt y/(2g))$.  
We can apply \re{trun} to rewrite \re{F(g)} as a Fredholm determinant 
\begin{align}\label{D-oper}
D_\ell(g) =  \det (1-{\bm K}_\ell\, \bm \chi)  \,.
\end{align}
We use below this relation to prove \re{Id}. 

The Bessel functions $\phi_n(x) = J_n(\sqrt x)$ satisfy a functional relation 
\begin{align} \label{Bes}
  2n{\phi_n(x)\over\sqrt x}=\phi_{n+1}(x)+\phi_{n-1}(x)\,.
\end{align}  
We apply this relation to verify that the kernel \re{K-sum} is related to the kernels $K_{\ell-1}(x,y)$ and $K_{\ell+1}(x,y)$
as
\begin{align}\notag\label{KK1} 
{}&K_\ell(x,y)  = \sqrt{x} K_{\ell+1}(x,y) {1\over\sqrt{y}}+\phi_\ell(x)\frac{\phi_{\ell+1}(y)}{2\sqrt{y}}
\\
{}& \phantom{K_\ell(x,y)} =\sqrt{x} K_{\ell-1}(x,y) {1\over\sqrt{y}} -\phi_\ell(x)\frac{\phi_{\ell-1}(y)}{2\sqrt{y}}\,.
\end{align}
Rewriting these relations in an operatorial form, we can obtain two equivalent representations of the Fredholm determinant \re{D-oper} 
\begin{align}\notag\label{D-two}
D_\ell = \det \lr{1-\sqrt{\bm x}\bm K_{\ell+1} {\bm \chi\over\sqrt{\bm x}} -\ket{\phi_\ell}\bra{ \phi_{\ell+1}}\frac{\bm \chi}{2\sqrt{\bm x}} }
\\
=\det \lr{1-\sqrt{\bm x}\bm K_{\ell-1} {\bm \chi\over\sqrt{\bm x}}+\ket{\phi_\ell}\bra{ \phi_{\ell-1}}\frac{\bm \chi}{2\sqrt{\bm x}} }
\end{align}
Using invariance of the determinant under a similarity transformation, we can simplify the first relation as
\begin{align}\notag\label{rel1}
D_\ell{}&=\det \lr{1- \bm K_{\ell+1}\bm \chi-\frac{1}{2\sqrt{\bm x}}\ket{\phi_\ell}\bra{ \phi_{\ell+1}} \bm \chi}
\\\notag
{}&=D_{\ell+1}\det \lr{1-(1- \bm K_{\ell+1}\bm \chi)^{-1}\frac{1}{2\sqrt{\bm x}}\ket{\phi_\ell}\bra{ \phi_{\ell+1}}\bm \chi }
\\
{}&=D_{\ell+1}\lr{1-\bra{ \phi_{\ell+1}} \bm \chi (1-\bm K_{\ell+1}\bm \chi)^{-1}\frac{1}{2\sqrt{\bm x}}\ket{\phi_\ell}}\,,
\end{align}
where in the second relation we took into account \re{D-oper}.

Repeating the same calculation using the second relation in \re{D-two}, we obtain the analogous relation between $D_\ell$ and $D_{\ell-1}$. Being combined together with \re{rel1} this leads to 
\begin{align}\notag\label{D/DD}
{D_\ell^2 \over D_{\ell+1}D_{\ell-1}} {}& = \lr{1-\bra{ \phi_{\ell+1}} \bm \chi (1- \bm K_{\ell+1} \bm \chi)^{-1}\frac{1}{2\sqrt{\bm x}}\ket{\phi_\ell}}
\\
{}& \times\lr{1+\bra{ \phi_{\ell-1}} \bm \chi (1- \bm K_{\ell-1}\bm \chi)^{-1}\frac{1}{2\sqrt{\bm x}}\ket{\phi_\ell}}.
\end{align}
Applying the identity \re{Bes} we can rewrite the right-hand side as
\begin{align}\notag
{}&1-{1\over 4\ell}\bra{ \phi_{\ell+1}} \bm \chi (1- \bm K_{\ell+1}\bm \chi)^{-1} \ket{\phi_{\ell+1}+\phi_{\ell-1}}
\\\notag
{}&+{1\over 4\ell}\bra{ \phi_{\ell-1}} \bm \chi (1- \bm K_{\ell-1}\bm \chi)^{-1} \ket{\phi_{\ell+1}+\phi_{\ell-1}}
\\
{}&-\frac14 \bra{ \phi_{\ell+1}} \bm \chi (1- \bm K_{\ell+1}\bm \chi)^{-1}\frac{1}{\sqrt{\bm x}}\ket{\phi_\ell}\bra{ \phi_{\ell}} \frac{1}{\sqrt{\bm x}} (1-\bm \chi \bm K_{\ell-1})^{-1}\bm \chi\ket{\phi_{\ell-1}}\,.
\end{align}
Expression on the last line involves the kernel ${\phi_\ell(x)\phi_{\ell}(y)/\sqrt{x y}}$, which can be futher simplified with a help of the identity
\begin{align}\label{K-diff}
{\phi_\ell(x)\phi_{\ell}(y) \over \sqrt{x y}}={1\over\ell}\lr{K_{\ell-1}(x,y)- K_{\ell+1}(x,y)}\,,
\end{align}
which follows from \re{KK1} and \re{Bes}. In this way, we obtain from \re{D/DD} 
\begin{align}\notag\label{rhs}
2\ell\left({D_\ell^2\over  D_{\ell-1}D_{\ell+1}} -1\right){}&={1\over 2}\bra{ \phi_{\ell-1}} \bm \chi (1- \bm K_{\ell-1}\bm \chi)^{-1} \ket{\phi_{\ell-1}}
\\
{}&-{1\over 2}\bra{ \phi_{\ell+1}}\bm  \chi (1- \bm K_{\ell+1}\bm \chi)^{-1} \ket{\phi_{\ell+1}}\,.
\end{align}
This yields the right-hand side of \re{Id}.

Let us examine the left-hand side of \re{Id}. We use the relations \re{D-oper} and \re{kernels} to get
\begin{align}\notag\label{lhs}
 g\partial_g \log {D_{\ell+1}\over D_{\ell-1}} {}&=  \tr\left({1\over 1- \bm K_{\ell-1} \bm \chi} \bm K_{\ell-1} g\partial_g \chi\right)- \tr\left({1\over 1- \bm K_{\ell+1} \bm \chi} \bm K_{\ell+1} g\partial_g \bm \chi\right)
\\\notag
{}&=-2\tr\left({1\over 1- \bm K_{\ell-1} \bm \chi} \bm K_{\ell-1} [\bm x\partial_{\bm x}, \bm \chi]\right)+2\tr\left({1\over 1- \bm K_{\ell+1} \bm \chi} \bm K_{\ell+1} [\bm x\partial_{\bm x}, \bm \chi]\right)
\\ 
{}&=2\tr\left({1\over 1- \bm K_{\ell-1} \bm \chi} [\bm x\partial_{\bm x},\bm K_{\ell-1}]\chi\right)-2\tr\left({1\over 1- \bm K_{\ell+1} \bm \chi} [\bm x\partial_{\bm x},\bm K_{\ell+1}] \bm \chi\right),
\end{align}
where in the second relation we used the identity  
\begin{align}
\lr{g\partial_g+2 x\partial_x}\chi\lr{\sqrt x\over 2g}=0\,.
\end{align}
The commutator $[\bm x\partial_{\bm x},\bm K_{\ell}]$ on the last line of \re{lhs} has the kernel $(x\partial_x + y\partial_y +1)K_{\ell}(x,y)$. Using  \re{K-sum} we find that it is given by $\phi_\ell(x)\phi_\ell(y)/4$, leading to
\begin{align}
[\bm x\partial_{\bm x},\bm K_{\ell}]=\frac14\ket{\phi_\ell}\bra{\phi_\ell}\,.
\end{align}
Substituting this relation into \re{lhs} we find that the right-hand side of \re{rhs} and \re{lhs} coincide. 
 This completes the proof of \re{Id}.
 
\section{Sum rules}\label{app:sum}

In this appendix, we derive the sum rule \re{sum-rule} for the expansion coefficients appearing in \re{ansatz}. Recall that the function $d_n(g)$, defined in \re{ansatz}, satisfies the differential equations \re{eqs} for arbitrary functions $f_+(g)$ and $f_-(g)$.  The expansion coefficients in \re{ansatz} are independent of the choice of these functions as well as of the parameters $\ell$ and $g$. 

We can take advantage of this property to choose $\ell=1$ and  
\begin{align}\label{spec}
d\log f_+ = d\big(g^{h_+}\big)\,, \qquad\qquad d\log f_- = d\big(g^{h_-}\big)\,,
\end{align}
where $h_+$ and $h_-$ are arbitrary parameters. 
In this case, a general expression  \re{ansatz}  takes the form
\begin{align}\label{d-ell=1} 
d_n(g)=\sum c_{\sigma_1\sigma_2\cdots\sigma_k} I_{\sigma_1\sigma_2\cdots\sigma_k}(g) \,,
\end{align}
where $k=n(n-1)/2$.  
Furthermore, substituting \re{spec} into \re{chen}, we find the iterated integrals in a close form  
\begin{align}\label{I-h}
I_{\sigma_1\sigma_2\cdots \sigma_k} = {h_{\sigma_1}\over \sum_{i=1}^k  h_{\sigma_i}} \dots   {h_{\sigma_{k-1}}\over \sum_{i=k-1}^k h_{\sigma_i}} g^{\sum_{i=1}^k  h_{\sigma_i}}
=g^{k_+ h_++ k_- h_-}\prod_{p=1}^{k-1} {h_{\sigma_p}\over \sum_{i=p}^k  h_{\sigma_i}}  
\end{align}
where $k_\pm$ denotes the total number of `$\pm$' in the sequence $(\sigma_1\sigma_2\dots \sigma_k)$ (with $\sigma_i=\pm$). 

\subsection*{Exact solution}

For the functions $f_\pm(g)$ defined in \re{spec} the differential equations \re{eqs} can be easily solved by using the following ansatz
\begin{align}\label{ans2}
d_{2k}(g) = A_k\, g^{2\alpha_k}\,, \qquad\qquad d_{2k-1}(g) = B_k\, g^{2\beta_k}\,,
\end{align}
and choosing the parameters $h_\pm$ in \re{spec} as
\begin{align}
h_+=2z\,,\qqqquad h_-=2(1-z)\,,
\end{align}
where $z$ is arbitrary.
Inserting it into \re{eqs} leads to a system of coupled recurrence relations for the exponents $\alpha_k$, $\beta_k$, and the normalization factors $A_k$, $B_k$. These relations are supplemented by the initial conditions $\alpha_0 = \beta_1 = 0$ and $A_0 = B_1 = 1$, which follow from \re{ini}.
Solving the recurrence relations, we obtain
\begin{align}\notag\label{AB}
\alpha_k &= k(k + z - 1)\,, \\[4mm] \notag
\beta_k  &= (k - 1)(k + z - 1)\,, \\[2mm] \notag
A_k &= \frac{(1 - z)^{k(k - 1)}\, z^{(k - 1)^2}}{(1 + z)^{2k - 2}}   
\frac{G (2 + z)}{G(2k + z)}\Gamma^{2k - 2}(2 + z)\,, \\[2mm]
B_k &= \frac{(1 - z)^{(k - 1)^2}\, z^{(k - 1)(k - 2)}}{(1 + z)^{2k - 3}}   
\frac{G(2 + z)}{G(2k - 1 + z)}\Gamma^{2k - 3}(2 + z)\,,
\end{align}
where $G(x)$ denotes the Barnes $G$-function.

Replacing $d_n(g)$ on the left-hand side of \re{d-ell=1} with its expression \re{ans2} and \re{AB} we obtain a nontrivial relation for the expansion coefficients that is valid for arbitrary $z$. This relation simplifies significantly at $z=1/2$. In this case, we have from \re{ans2} and \re{I-h}
\begin{align}\notag
{}& d_n  \Big|_{z=1/2} = (g/ 2)^{n(n-1)/2} {\pi^{n/2}G(\ft12)\over G(n+\ft12)}\,,
\\[2mm]
{}& I_{\sigma_1\sigma_2\dots \sigma_k} \Big|_{z=1/2} = {g^{k}\over k!}\,.
\end{align}
Substituting these relations into \re{d-ell=1} and using \re{k-n}, we observe that both sides of the equation are proportional to 
$g^{n(n-1)/2}$. Matching the coefficients then leads to the sum rule \re{sum-rule}.

Additional sum rules for the coefficients $ c_{\sigma_1 \sigma_2 \dots \sigma_k} $ can be obtained by expanding both sides of \re{d-ell=1} around $ z =1/2$. We have verified that these sum rules are sufficient to determine all the coefficients in the expansion of $ d_n $ for $ n \leq 4 $. However, starting from $ n = 5 $, some coefficients remain undetermined.

\subsection*{Paths weighted by height}

We demonstrate in Section~\ref{sect:paths} that, for arbitrary $I-$function, the relation \re{d-ell=1} can be interpreted as a sum over paths on the square lattice. Furthermore, we show below that the 
$I-$functions defined in \re{I-h} admit a simple interpretation in terms of the height function associated with these paths.

\begin{figure}[t]
\begin{centering}
\includegraphics[width=0.95\textwidth]{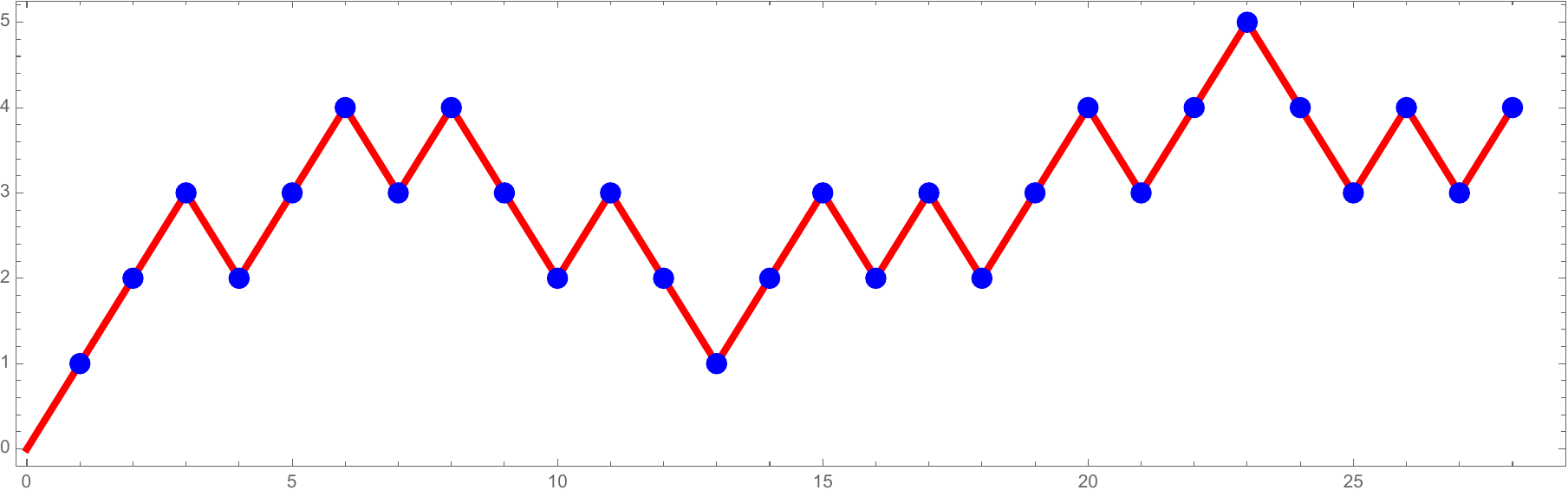}
\par\end{centering}
\caption{The path (`mountain ridge') contributing to the $W-$function defined in \re{H}. It is obtained by reflecting the path shown in Figure~\ref{red-path} about the origin. The blue dots indicate the endpoints of the up- and down-steps. For 
$h_+=-h_-=1$, the $W-$function is given by the product of the inverse heights at these blue dots.}
\label{reverse-path}
\end{figure}

Let us define an auxiliary weight function  
\begin{align}\label{H}
W_{\sigma_1\sigma_2\cdots\sigma_k} ={1\over h_{\sigma_1}} {1\over h_{\sigma_1}+h_{\sigma_2}}\dots{1\over h_{\sigma_1}+\dots+h_{\sigma_k}}\,.
\end{align}
It is related to the function \re{I-h} as 
\begin{align}\label{I-height}
I_{\sigma_1\sigma_2\cdots \sigma_k} =  (h_+ g^{h_+})^{k_+} (h_- g^{h_-})^{k_-}W_{\sigma_k\cdots\sigma_2\sigma_1}\,.
\end{align}
Note that the $W-$function involves the sequence in reverse order compared to the sequence in the $I-$function on the left-hand side. Consequently, the path corresponding to the $W-$function can be obtained by reflecting the $I-$function's path about the origin. For example, the path shown in Figure~\ref{red-path} gets transformed into the path depicted in Figure~\ref{reverse-path}. 

The paths contributing to the function \re{H} consist of $k = n(n-1)/2$ segments, comprising $k_+$ up-steps (\( \nearrow \)) and \( k_- \) down-steps (\( \searrow \)). We assign a height to each step of the path. The first step, originating at the origin, is assigned a height of \( h_+ \). For each subsequent step, the height increases by \( h_+ \) following an up-step and decreases by \( h_- \) after a down-step.\,\footnote{For the purpose of this discussion, we assume \( h_- < 0 \).} The weight function  \re{H} is then given by the product of the inverse heights of all steps along the path. The example shown in Figure~\ref{reverse-path} corresponds to the special case where \( h_+ = -h_- = 1 \).

\section{Flux tube correlators}\label{app:flux} 
 
For the symbol function $\chi_{\text{ft}}(x)$ defined in \re{chis}, the determinant \re{F(g)} can be computed  for the lowest values of $\ell=0,1,2$ in terms of hyperbolic functions, see \re{Dft-spec}. In this appendix, we apply \re{d-def} and \re{ansatz} to compute the function $D_{\rm ft,\ell}(g)$ for $\ell\ge 3$.
 
For $\ell=1$ and $n=1$ the relations \re{d-def} look as
\begin{align}\notag\label{DD}
{}& {D_2(g)\over D_0(g)} = I_+(g)/g^2\,,
\\[2mm]
{}& {D_3(g)\over D_1(g)}= 4  I_{-++}(g)/g^4\,.
\end{align}
To compute the iterated integrals $I_+(g)$ and $I_{-++}(g)$, it is convenient to introduce the new variable
\begin{align}\label{z-g}
z=e^{-4\pi g}
\end{align} 
and rewrite the first two relations in \re{Dft-spec} in terms of $z$
\begin{align} \notag 
{}& D_0= \left[\frac{(1+z)^3 \log (1/z)}{8 (1-z) z} \right]^{1/8}
\,, 
\\
{}& D_1= \left[\frac{2 (1-z)^3}{z (1+z) \log ^3(1/z)}\right]^{1/8}\,.
\end{align}
Inserting these expressions into \re{f+-} and setting $\ell=1$, we obtain
\begin{align} \notag
{}& (\log f_+(g))'= {1\over\pi} \frac{1-z}{1+z}\,, 
\\
{}& (\log f_-(g))'= 4\pi  \frac{1+z}{1-z} \,.
\end{align}
At the next step, we apply \re{chen} and compute the iterated integral in the first relation in \re{DD}
\begin{align}
 I_+(g) =   {1\over (2\pi)^2} \int^1_z  {dz\over z} \frac{1-z}{1+z}= {1\over 2\pi^2}\log\lr{1+z\over 2\sqrt z}
={1\over 2\pi^2}\log \cosh(2\pi g)\,.
\end{align} 
Substituting this relation into \re{DD} we obtain 
\begin{align}
{D_2(g)\over D_0(g)}={1\over 2(\pi g)^2}\log \cosh(2\pi g)
\end{align}
in agreement with \re{Dft-spec}.

The integral in the second relation in \re{DD} admits two equivalent representations
\begin{align}\notag\label{I-dlog}
I_{-++}(g){}&={1\over (2\pi)^4}\int^1_z  {dz_1\over z_1} \frac{1+z_1}{1-z_1} \int^1_{z_1}  {dz_2\over z_2} \frac{1-z_2}{1+z_2}  \int^1_{z_2}  {dz_3\over z_3} \frac{1-z_3}{1+z_3}
\\[2mm]
{}&={1 \over 2\pi^4}\int^z_1  d \log\lr{1- z_1\over 2\sqrt z_1}\int^{z_1}_1 d \log\lr{1+z_2\over 2\sqrt z_2}\int^{z_{2}}_1 d \log\lr{1+z_3\over 2\sqrt z_3}\,.
\end{align} 
The second relation is known as  \textit{d-log} representation.

The integral \re{I-dlog} can be expressed in terms of harmonic polylogarithms (HPL)
\begin{align}\notag\label{I-HPL}
I_{-++}(g)= {1\over 4\pi^4} {}&\bigg[\frac{1}{2} H_{0,0,-1}(z)-H_{0,-1,-1}(z)+\frac{1}{2} H_{0,-1,0}(z)+H_{1,0,-1}(z)-\frac{1}{4}
   H_{0,0,0}(z) 
\\{}&\notag   
  -2 H_{1,-1,-1}(z) +H_{1,-1,0}(z)-\frac{1}{2}H_{1,0,0}(z) -\frac{1}{2}H_{-1}(1) H_{0,0}(z)
\\[2mm]{}&   \notag
+2H_{-1}(1) H_{1,-1}(z)-H_{-1}(1)H_{1,0}(z)+H_{-1}(1) H_{0,-1}(z)
\\[2mm]{}& 
-H_{-1,-1}(1)H_{0}(z) -2 H_{-1,-1}(1)H_{1}(z)+2H_{-1,-1,1}(1)\bigg],
\end{align}
where $z$ is defined in \re{z-g}. The HPL functions are defined recursively as \cite{Remiddi:1999ew,Maitre:2005uu}
\begin{align}\notag\label{hpl}
{}& H_{a_1,a_2,a_3\dots}(z) = \int_0^z dt f_{a_1}(t) H_{a_2,a_3,\dots}(t)\,,
\\[2mm]
{}& f_1(x)={1\over 1-x}\,,\qquad f_0(z)={1\over z}\,,\qquad f_{-1}(z)={1\over 1+z}\,.
\end{align}
Note that the integrals in \re{I-dlog} and \re{hpl} differ in their lower limits of integration. This explains the presence of $H_{a_1,a_2,a_3\dots}(1)$ in the relation \re{I-HPL}. 

Combining together the relations \re{I-HPL} and \re{DD} we can compute the function $D_3(g)$. 
In distinction with \re{Dft-spec}, this function is given by a multi-linear combination of HPL's functions.  
The same property holds for the functions $D_n(g)$ for $n>3$. Applying \re{d-def} and \re{ansatz}, we can use the obtained expressions for $d_n(g)$ (see \re{sol1}, \re{d4}, \re{sol2} and \re{sol3}) to express $D_n(g)$ as a linear combination of the iterated integrals \re{chen} of weight $k=n(n-1)/2$. In a close analogy with \re{I-dlog}, these integrals can be written in the \textit{d-log} form
\begin{align}
I_{\sigma_1\sigma_2\cdots\sigma_k} ={2^{k_-} \over (2\pi^2)^{k_+}}\int^z_1  d \log\lr{1+\sigma_1 z_1\over 2\sqrt z_1}\int^{z_1}_1 d \log\lr{1+\sigma_2 z_2\over 2\sqrt z_2}... \int^{z_{n-1}}_1 d \log\lr{1+\sigma_k z_n\over 2\sqrt z_n}
\end{align}
where $k_+$ and $k_-$ is the total number of pluses and minuses in the sequence $(\sigma_1\cdots\sigma_k)$. Similar to \re{I-HPL}, these integrals can be 
expanded as multi-linear combinations of HPL's $H_{a_1a_2\dots}(z)$ and $H_{a_1a_2\dots}(1)$ with indices $a_i\in \{-1,0,1\}$. 
 
 \section{Solution for $d_6$}\label{app:d6}
 
Solving the equations \re{eqs}, we obtained the following result 
\footnotesize
 \begin{align}\notag\label{sol3}
{}& d_6(g) = 64 
\\\notag {}& \times 
 \Big( 30 I_{+--+-+-+--+++++}+18 I_{+--+-+-+-+-++++} 
+9
   I_{+--+-+-+-++-+++}+3 I_{+--+-+-+-+++-++} \\ \notag {}&  
   +24
   I_{+--+-+-++--++++}+12 I_{+--+-+-++-+-+++}  
   +4
   I_{+--+-+-++-++-++}+9 I_{+--+-+-+++--+++}   \\ \notag {}&
   +3
   I_{+--+-+-+++-+-++}+90 I_{+--+-++---+++++}  
   +54
   I_{+--+-++--+-++++}+27 I_{+--+-++--++-+++}   \\ \notag {}&
   +9
   I_{+--+-++--+++-++}+42 I_{+--+-++-+--++++}   
   +21
   I_{+--+-++-+-+-+++}+7 I_{+--+-++-+-++-++}   \\ \notag {}&
   +12
   I_{+--+-++-++--+++}+4 I_{+--+-++-++-+-++}  
   +54
   I_{+--+-+++---++++}+27 I_{+--+-+++--+-+++}    \\ \notag {}&
   +9
   I_{+--+-+++--++-++}+9 I_{+--+-+++-+--+++}    
   +3
   I_{+--+-+++-+-+-++}+90 I_{+--++--+--+++++}    \\ \notag {}&
   +54
   I_{+--++--+-+-++++}+27 I_{+--++--+-++-+++}   
   +9
   I_{+--++--+-+++-++}+72 I_{+--++--++--++++}    \\ \notag {}&
   +36
   I_{+--++--++-+-+++}+12 I_{+--++--++-++-++}   
   +27
   I_{+--++--+++--+++}+9 I_{+--++--+++-+-++}    \\ \notag {}&
   +270
   I_{+--++-+---+++++}+162 I_{+--++-+--+-++++}    
   +81
   I_{+--++-+--++-+++}+27 I_{+--++-+--+++-++}    \\ \notag {}&
   +96
   I_{+--++-+-+--++++}+48 I_{+--++-+-+-+-+++}    
   +16
   I_{+--++-+-+-++-++}+21 I_{+--++-+-++--+++}    \\ \notag {}&
   +7
   I_{+--++-+-++-+-++}+72 I_{+--++-++---++++}    
   +36
   I_{+--++-++--+-+++}+12 I_{+--++-++--++-++}    \\ \notag {}&
   +12
   I_{+--++-++-+--+++}+4 I_{+--++-++-+-+-++}    
   +540
   I_{+--+++----+++++}+324 I_{+--+++---+-++++}    \\ \notag {}&
   +162
   I_{+--+++---++-+++}+54 I_{+--+++---+++-++}    
   +162
   I_{+--+++--+--++++}+81 I_{+--+++--+-+-+++}    \\ \notag {}&
   +27
   I_{+--+++--+-++-++}+27 I_{+--+++--++--+++}   
   +9
   I_{+--+++--++-+-++}+54 I_{+--+++-+---++++}    \\ \notag {}&
   +27
   I_{+--+++-+--+-+++}+9 I_{+--+++-+--++-++}   
   +9
   I_{+--+++-+-+--+++}+3 I_{+--+++-+-+-+-++}    \\ \notag {}&
   +30
   I_{+-+--+-+--+++++}+18 I_{+-+--+-+-+-++++}    
   +9
   I_{+-+--+-+-++-+++}+3 I_{+-+--+-+-+++-++}    \\ \notag {}&
   +24
   I_{+-+--+-++--++++}+12 I_{+-+--+-++-+-+++}    
   +4
   I_{+-+--+-++-++-++}+9 I_{+-+--+-+++--+++}     \\ \notag {}&
   +3
   I_{+-+--+-+++-+-++}+90 I_{+-+--++---+++++}    
   +54
   I_{+-+--++--+-++++}+27 I_{+-+--++--++-+++}    \\ \notag {}&
   +9
   I_{+-+--++--+++-++}+42 I_{+-+--++-+--++++}    
   +21
   I_{+-+--++-+-+-+++}+7 I_{+-+--++-+-++-++}    \\ \notag {}&
   +12
   I_{+-+--++-++--+++}+4 I_{+-+--++-++-+-++}    
   +54
   I_{+-+--+++---++++}+27 I_{+-+--+++--+-+++}    \\ \notag {}&
   +9
   I_{+-+--+++--++-++}+9 I_{+-+--+++-+--+++}    
   +3
   I_{+-+--+++-+-+-++}+30 I_{+-+-+--+--+++++}    \\ \notag {}&
   +18
   I_{+-+-+--+-+-++++}+9 I_{+-+-+--+-++-+++}    
   +3
   I_{+-+-+--+-+++-++}+24 I_{+-+-+--++--++++}    \\ \notag {}&
   +12
   I_{+-+-+--++-+-+++}+4 I_{+-+-+--++-++-++}    
 +9
   I_{+-+-+--+++--+++}+3 I_{+-+-+--+++-+-++}    \\ \notag {}&
   +90
   I_{+-+-+-+---+++++}+54 I_{+-+-+-+--+-++++}    
   +27
   I_{+-+-+-+--++-+++}+9 I_{+-+-+-+--+++-++}    \\ \notag {}&
   +36
   I_{+-+-+-+-+--++++}+18 I_{+-+-+-+-+-+-+++}    
   +6
   I_{+-+-+-+-+-++-++}+9 I_{+-+-+-+-++--+++}    \\ \notag {}&
   +3
   I_{+-+-+-+-++-+-++}+36 I_{+-+-+-++---++++}    
   +18
   I_{+-+-+-++--+-+++}+6 I_{+-+-+-++--++-++}    \\ \notag {}&
   +6
   I_{+-+-+-++-+--+++}+2 I_{+-+-+-++-+-+-++}    
   +180
   I_{+-+-++----+++++}+108 I_{+-+-++---+-++++}    \\ \notag {}&
   +54
   I_{+-+-++---++-+++}+18 I_{+-+-++---+++-++}    
   +54
   I_{+-+-++--+--++++}+27 I_{+-+-++--+-+-+++}    \\ \notag {}&
   +9
   I_{+-+-++--+-++-++}+9 I_{+-+-++--++--+++}    
   +3
   I_{+-+-++--++-+-++}+18 I_{+-+-++-+---++++}    \\   {}&
   +9
   I_{+-+-++-+--+-+++}+3 I_{+-+-++-+--++-++}    
   +3
   I_{+-+-++-+-+--+++}+I_{+-+-++-+-+-+-++}\Big),
\end{align}
\normalsize
where the $I-$functions are given by \re{chen}.

Note that all expansion coefficients in \re{sol3} are positive integers. The maximal and minimal coefficients, equal to \( 540 \) and \( 1 \), respectively, multiply the functions \( I_{+--+++----+++++} \) and \( I_{+-+-++-+-+-+-++} \), which correspond to the lattice paths shown in Figures~\ref{max-path} and \ref{min-path}.

\bibliographystyle{JHEP}
\bibliography{papers}

\end{document}